\documentclass[preprint,3p,12pt,numbers,sort&compress]{elsarticle}

\usepackage{algorithm}
\usepackage{algpseudocode}
\usepackage{amssymb}
\usepackage{booktabs}
\usepackage{tablefootnote} 
\usepackage{amsmath}
\usepackage[utf8]{inputenc}
\usepackage{bm}
\usepackage[tbtags]{mathtools}
\usepackage{calc}

\usepackage{stackengine}
\usepackage{scalerel}
\usepackage{verbatimbox}
\usepackage[tbtags]{mathtools}
\usepackage[T1]{fontenc}
\usepackage[utf8]{inputenc}
\usepackage{lmodern}
\usepackage{combelow}
\usepackage{graphicx}
\usepackage[colorlinks=true, allcolors=blue]{hyperref}
\usepackage{soul,color}
\usepackage{amsmath}
\usepackage{amsthm}
\usepackage{eucal}
\usepackage{amssymb}
\usepackage{mathrsfs}
\usepackage{amsfonts}
\usepackage{caption}
\usepackage{subcaption}
\usepackage{miller}
\usepackage{url}
\usepackage{hyperref}
\usepackage{float}
\usepackage{multirow}

\journal{Journal}

\begin{document}
	
\begin{frontmatter}
		
\title{Generative diffusion modeling protocols for improving the Kikuchi pattern indexing in electron back-scatter diffraction}

\author[mech]{Meghraj Prajapat}
\affiliation[mech]{Department of Mechanical Engineering, organization={Indian Institute of Technology Bombay, Powai},
            city={Mumbai 400076},
            country={India}}
\author[mech,cminds]{Alankar Alankar \corref{cor}}
\affiliation[cminds]{Center for Machine Intelligence and Data Science (CMInDS), organization={Indian Institute of Technology Bombay, Powai},
            city={Mumbai 400076},
            country={India}}

\cortext[cor]{Corresponding author: Email: alankar.alankar@iitb.ac.in, Tel.: +91-9769415356,\\ Fax: +91-22-25726875}
\begin{abstract}
Electron back-scatter diffraction (EBSD) has traditionally relied upon methods such as the Hough transform and dictionary Indexing to interpret diffraction patterns and extract crystallographic orientation. However, these methods encounter significant limitations, particularly when operating at high scanning speeds, where the exposure time per pattern is decreased beyond the operating sensitivity of CCD camera. Hence the signal to noise ratio decreases for the observed pattern which makes the pattern noisy, leading to reduced indexing accuracy. This research work aims to develop generative machine learning models for the post--processing or on-the-fly processing of Kikuchi patterns which are capable of restoring noisy EBSD patterns obtained at high scan speeds. These restored patterns can be used for the determination of crystal orientations to provide reliable indexing results. We compare the performance of such generative models in enhancing the quality of patterns captured at short exposure times (high scan speeds). An interesting observation is that the methodology is not data-hungry as typical machine learning methods.
\end{abstract}
\end{frontmatter}
	
\section{Introduction}
\label{sec:intro}
\noindent In recent years, there have been significant advancements in hardware technology for electron microscopy, particularly for capturing crystallographic orientations using Electron Back-scatter Diffraction (EBSD). High-sensitivity CCD and CMOS cameras have been developed, enabling the capture of high-quality Kikuchi patterns even at shorter exposure times. While hardware improvements are crucial, the role of software in post-processing these measurements to generate meaningful results is equally important. Noteworthy is that good quality patterns also need a rigorous specimen preparation.

Traditionally, the Hough transform has been the primary method for processing Kikuchi patterns. This image processing technique converts the pattern into Hough space, allowing the identification of Kikuchi diffraction bands by detecting linear features in the transformed space. Based on the angles between these lines, the corresponding crystallographic planes can be inferred, since the relation between the lines and the crystallographic planes are well known and can be numerically determined. However, the accuracy of the Hough transform is highly dependent on the clarity of the Kikuchi bands, which diminishes significantly when scan speeds increase (exposure times drop) beyond the sensitivity of cameras \cite{schwartz2009electron}. Under such conditions, the observation is corrupted by undesired noise because of a lack of proper capturing of the signal. Hence, patterns become noisy and unclear, making the Hough transform unreliable. Other sources of noise and corruption of the EBSD patterns are due to improper specimen preparation and samples with a high degree of cold work \cite{krishna2023machine}.

A promising alternative to the Hough transform for indexing noisy, high-speed captured EBSD patterns is the dictionary indexing (DI) method \cite{chen2015dictionary}. In DI, the experimental pattern is compared with a pre-generated dictionary of patterns of reference material and the closest match is identified. Although it is robust against the noise introduced by high-speed scans, dictionary indexing has its own limitations. Its high computational complexity makes it unsuitable for real-time online indexing of patterns and as the size of the dictionary grows, especially for the materials with multiple phases, the search time increases significantly \cite{chen2015dictionary} and requires large offline storage. Furthermore, as pattern quality degrades due to other factors where search-based methods fail, the reliability of the DI method diminishes.

Another method is Spherical Indexing (SI) \cite{lenthe2019spherical}, which is currently one of the most advanced methods for EBSD orientation indexing. It uses spherical functions, enabling faster and more precise analysis compared to the dictionary indexing method. SI also relies on several preprocessing steps before indexing, which can significantly affect the results. These steps often require expert knowledge, and optimal procedures are sometimes identified through trial and error. Although SI is capable of indexing highly noisy patterns, its reliability becomes questionable when Kikuchi bands are not visually distinguishable. In such cases, restoring EBSD patterns prior to indexing and then applying a more robust and faster method like the Hough transform offers a more reliable approach for indexing extremely noisy Kikuchi patterns.

Recently, rapid advancements in machine learning (ML) and deep learning algorithms have been made that enable their applications in various engineering, automation and optimization tasks including materials science and characterization \cite{ostormujof2022deep, kaufmann2021efficient, kaufmann2020deep, stoll2021machine, choudhary2022recent}. Ostormujof et al \cite{ostormujof2022deep} used a Convolutional Neural Network (CNN) based U-Net architecture for the automated phase segmentation in EBSD maps for dual-phase steels. Kaufmann et al \cite{kaufmann2020deep} employed a deep neural network-based model for high-throughput automated space group classification. Kaufmann et al \cite{kaufmann2021efficient} further developed and optimized the method using a few shot transfer learning strategy. Andrews et al \cite{andrews2023denoising} developed a denoising autoencoder and trained on simulated patterns generated from high-quality orientation data to enhance the quality of Kikuchi patterns obtained on poorly prepared and defect-ridden surfaces. CNN-based orientation indexing has demonstrated superior performance as compared to traditional Hough transform-based methods, particularly for noisy patterns \cite{shen2019convolutional, ding2020indexing}. While the Hough-based approach becomes unreliable at higher noise levels specifically for patterns obtained at short exposure times, a well-trained CNN model can provide more reliable indexing results. These models are typically trained using both simulated and experimental data, incorporating observations across various noise levels. For training the CNN model misorientation between true and predicted orientations have been used as the loss function \cite{shen2019convolutional}. For further optimization of results, a better loss function named as disorientation loss function has been used which also takes into account the crystal symmetry \cite{ding2020indexing}.

Although CNN-based indexing offers reliable results even in the presence of significant noise, its effectiveness diminishes at extremely short exposure times, \cite{ding2020indexing} 
where key features of the pattern become indistinct. In our work, we have performed a case study of training the CNN model for orientation indexing and validating its robustness against noise due to high scan speeds. The intention of this study is to identify the level of noise based on exposure time per pattern, for which the CNN based orientation indexing fails to perform well \cite{ding2020indexing}. Based on our analysis, we recommend advanced methods to handle such high levels of noise in the EBSD patterns.  

\par Another class of models called generative deep learning models, have been applied for the parametric simulation (generation) of Kikuchi patterns \cite{Ding2023}. Generative models are a class of ML-based models that can generate new data that resembles the original dataset. These models aim to capture the underlying data distribution, allowing them to synthesize new instances from the learned distribution and have been widely used in fields such as image synthesis, text generation, and speech processing \cite{goodfellow2014generative, kingma2013auto, ho2020denoising}. The core idea behind generative models is to approximate the probability distribution of data points, either explicitly or implicitly, and sample from this distribution to create new data points. This capability makes generative models highly effective for applications requiring data augmentation, restoration, and synthesis \cite{wang2023review, saharia2022image}. 

\par There are several types of generative model with a distinct approach to data generation. Common types include Variational Autoencoders (VAEs) \cite{kingma2013auto}, Generative Adversarial Networks (GANs), introduced by Goodfellow et al. \cite{goodfellow2014generative} and Diffusion models \cite{ho2020denoising}. These models can simulate kinematic EBSD patterns by varying different simulation parameters \cite{Ding2023} similar to that in an experiment. GANs have also been applied for grain size quantification through EBSD maps \cite{anantatamukala2023generative}. Additionally, conditional GANs \cite{krishna2023machine} have been utilized for the denoising of the EBSD patterns. Such noise may occur due to small exposure times in EBSD analyses, surface roughness or insufficient specimen surface preparation for EBSD analyses.

GANs have been remarkably successful in generating high-resolution images, artistic creations and image-to-image translation \cite{shahriar2022gan, ledig2017photo, wang2018esrgan}. However, they present several challenges, including mode collapse \cite{zhang2018convergence}, where the generator produces limited variations in data, and training instability \cite{zhang2018convergence}, which requires careful balancing of the two networks, i.e, generator and discriminator. GANs and diffusion models both model the probablity distributions implicitly not explicitly , but diffusion model uses likelihood based training objective, with iterative denoising framework, making them more stable. Traditional GANs do not explicitly model probability distributions through likelihood-based objective but instead rely on an adversarial loss, where the generator is trained to fool the discriminator in a minimax game.

\par Diffusion models, a more recent class of generative models, take a fundamentally different approach by learning to reverse a noise corruption process to recover a clean image using a likelihood-based training through variational bounds \cite{ho2020denoising}. Diffusion models have demonstrated superior performance in high-quality image restoration tasks such as denoising, super-resolution, and inpainting \cite{saharia2022image, rombach2022high, kawar2022denoising} surpassing GANs in generating realistic images with better mode coverage and stability. The effectiveness of diffusion models comes from their iterative denoising framework, which follows a Markovian process \cite{ho2020denoising}. This approach of progressively refining the images allows them to capture fine details with greater stability and diversity compared to GANs, which also makes them more generalised and less prone to overfit. However, this iterative nature leads to long sampling times, making them computationally expensive. However, several advanced algorithms such as DDIM (Denoising Diffusion Implicit Models) \cite{song2020denoising}, LDM (Latent Diffusion Models) \cite{rombach2022high}, and EDM (Efficient Diffusion Models) \cite{ulhaq2022efficient} have been developed to address this challenge and to enhance sampling speed. Through these innovations, the sampling speed of diffusion models has improved significantly. These advancements enable faster image generation while maintaining high quality and diversity in the outputs.

Falaleev and Orlov \cite{falaleev2025self} reported a self-controlled diffusion model for denoising of EBSD patterns, which involves a feedback-based adaptive denoising framework. However, unlike conventional conditional generative diffusion models that work on probabilistic training objective, their approach follows a deterministic training framework. While it demonstrates strong performance in denoising EBSD patterns, its deterministic nature limits its generalisation compared to probabilistic frameworks based on Denoising Diffusion Probabilistic Models (DDPM). 

The capability of Generative diffusion models to produce high-quality and stable results makes them a compelling option for improving low-quality EBSD patterns. Instead of relying on indexing poor-quality patterns, the restoration of such patterns to their high-quality form prior to indexing offers a better, reliable approach for high-speed EBSD pattern analysis. The diffusion models, due to their wide mode coverage capabilities, are well-suited to handle diverse degradation types within a unified framework \cite{saharia2022palette, mandal2025unicorn}. This is particularly advantageous for EBSD data, where degradation may arise from various sources such as improper sample preparation, residual strain, or experimental inconsistencies.

\par Our diffusion-based restoration framework holds strong potential to serve as a universal solution for enhancing Kikuchi pattern quality across a wide spectrum of degradation sources and experimental configurations. With the rapid progress in artificial intelligence and machine learning, the future of these models in EBSD post-processing software is promising, offering the potential to push beyond the limits imposed by current hardware optimization. Applications of DDPM-based frameworks for the restoration of Kikuchi patterns achieved using high-speed EBSD scans yet remains unexplored. The aim of current research work is to explore the applications of these models for enhancement of EBSD measurements obtained at high scan speeds or short exposure times per patterns.



\section{Dataset details}

\begin{figure}[H]
    \centering
    \includegraphics[width=1\linewidth]{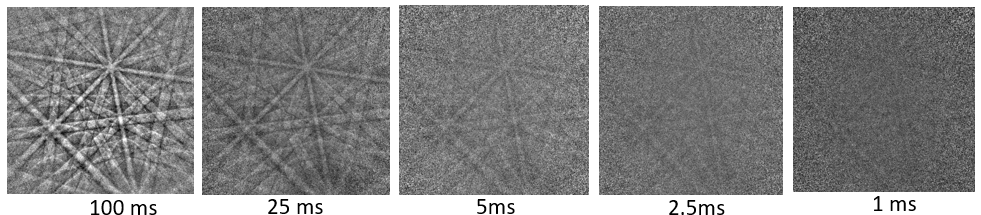}
    \caption{Quality of Kikuchi patterns as a function of exposure time. A 100 ms exposure time is sufficient for the EBSD camera to capture all the fine details like lines and poles of the Kikuchi pattern clearly. Signal to noise ratio decreases for decreasing exposure time.}
    \label{fig:speed vs quality}
\end{figure}

\noindent For this work, the EBSD data was received from ThermoFisher Inc. EBSD data was collected from a region of 100 $\mu m$ $\times$ 100 $\mu m$ area using a 50 $\times$ 50 pixel grid. Kikuchi patterns were captured by using a CCD camera. For each point, the signals are recorded for various exposure time e.g.  0.25 ms, 1 ms, 2 ms, up to a total duration of 25 ms. Such data collection allows us to obtain the Kikuchi patterns and therefore the orientation maps of various quality. 
As evident in Figure \ref{fig:speed vs quality}, for short exposure time i.e. higher scan speed, the pattern observed has a very low signal-to-noise ratio and the Kikuchi band are not very clearly visible. 

\section{CNN based surrogate models for orientation indexing of EBSD patterns}
\begin{figure}[!ht]
    \centering
    \includegraphics[width=1\linewidth]{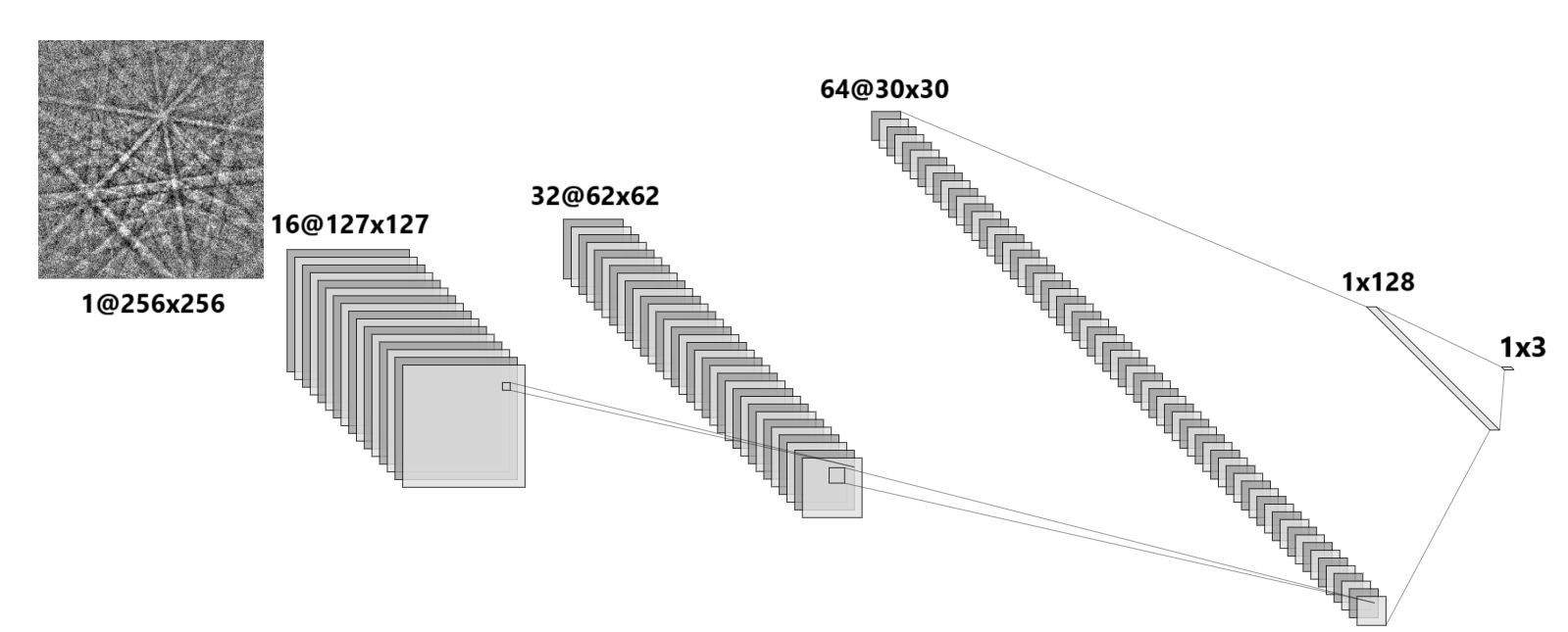}
    \caption{CNN architecture mapping 3-dimensional output (Euler angles) to the input image of Kikuchi pattern. The notation used to mention the architecture is `number of channels $@$ size of the features'.}
    \label{fig:KikchiCNN}
\end{figure}

\noindent Before we describe our exploration of diffusion-based generative models, we describe how CNN based model can provide a quick, however limited work around. A CNN model can serve as a potential surrogate for the EBSD orientation indexing. Instead of manually extracting Kikuchi bands and performing Hough peak detection, a well-trained CNN can directly learn to map an input Kikuchi pattern to its corresponding crystallographic orientation in form of Euler angles \cite{ding2020indexing, Ding2023, shen2019convolutional}. CNNs consist of multiple convolutional layers, which apply filters (or kernels) across the entire Kikuchi pattern. These filters help the CNN learn local features in the pattern, such as the edges, lines, and orientations of Kikuchi bands. Each filter extracts a particular feature of the pattern by scanning across the image and performing element-wise multiplications with small regions of the image (called receptive fields). A simplified architecture of CNN model is shown in Figure \ref{fig:KikchiCNN}. The trained CNN model takes noisy images of Kikuchi patterns and predicts Euler angles which are used in MTex \cite{bachmann2010texture} for creating the orientation map.

\vspace{1em}



\subsection{Predictive orientation indexing and exposure time trade-off}
\begin{figure}[H]
    \centering
    \begin{subfigure}{0.32\linewidth}
        \centering
        \includegraphics[width=\linewidth]{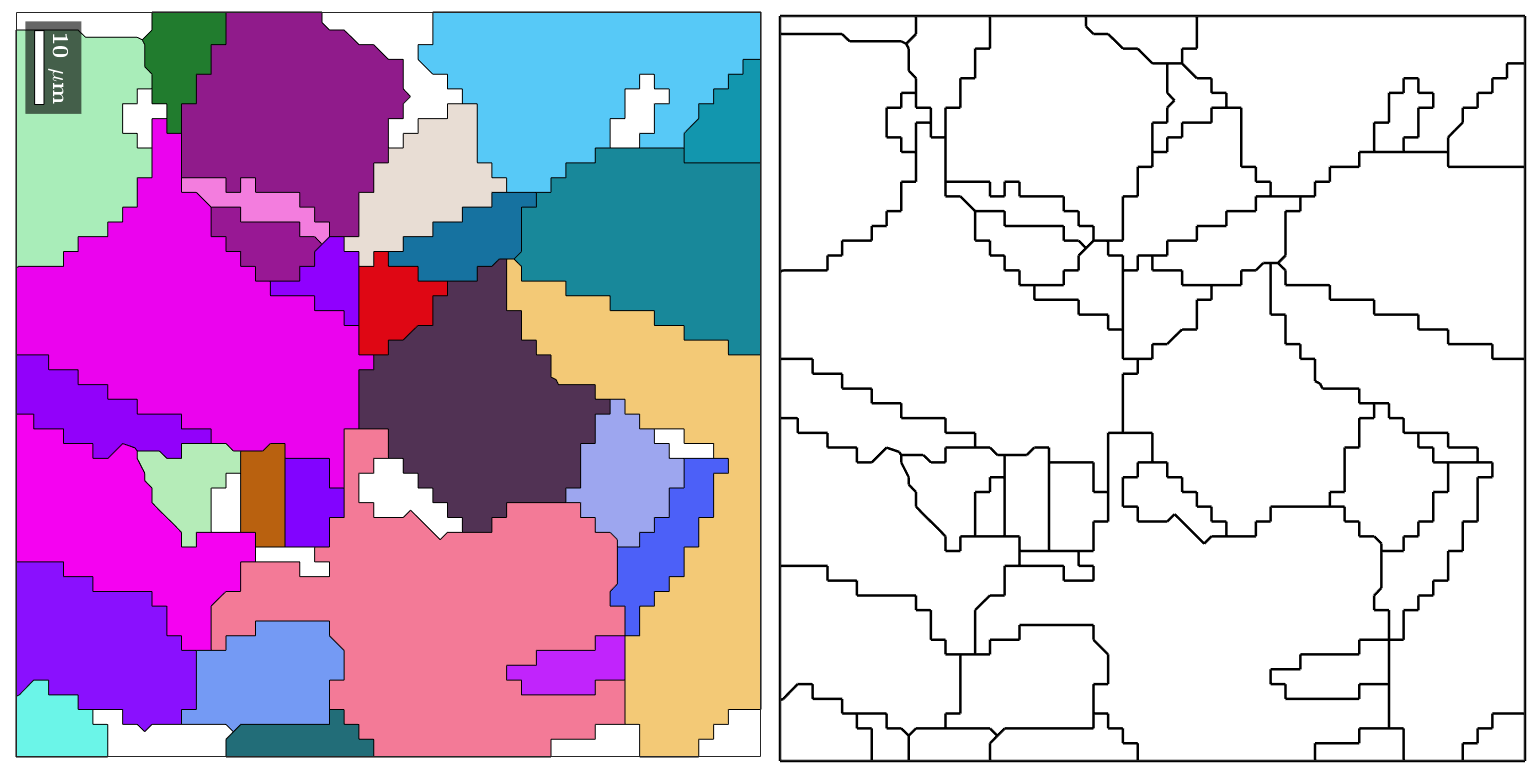}
        \caption{}
        \label{fig:true_gb}
    \end{subfigure}
    \hfill
    \begin{subfigure}{0.32\linewidth}
        \centering
        \includegraphics[width=\linewidth]{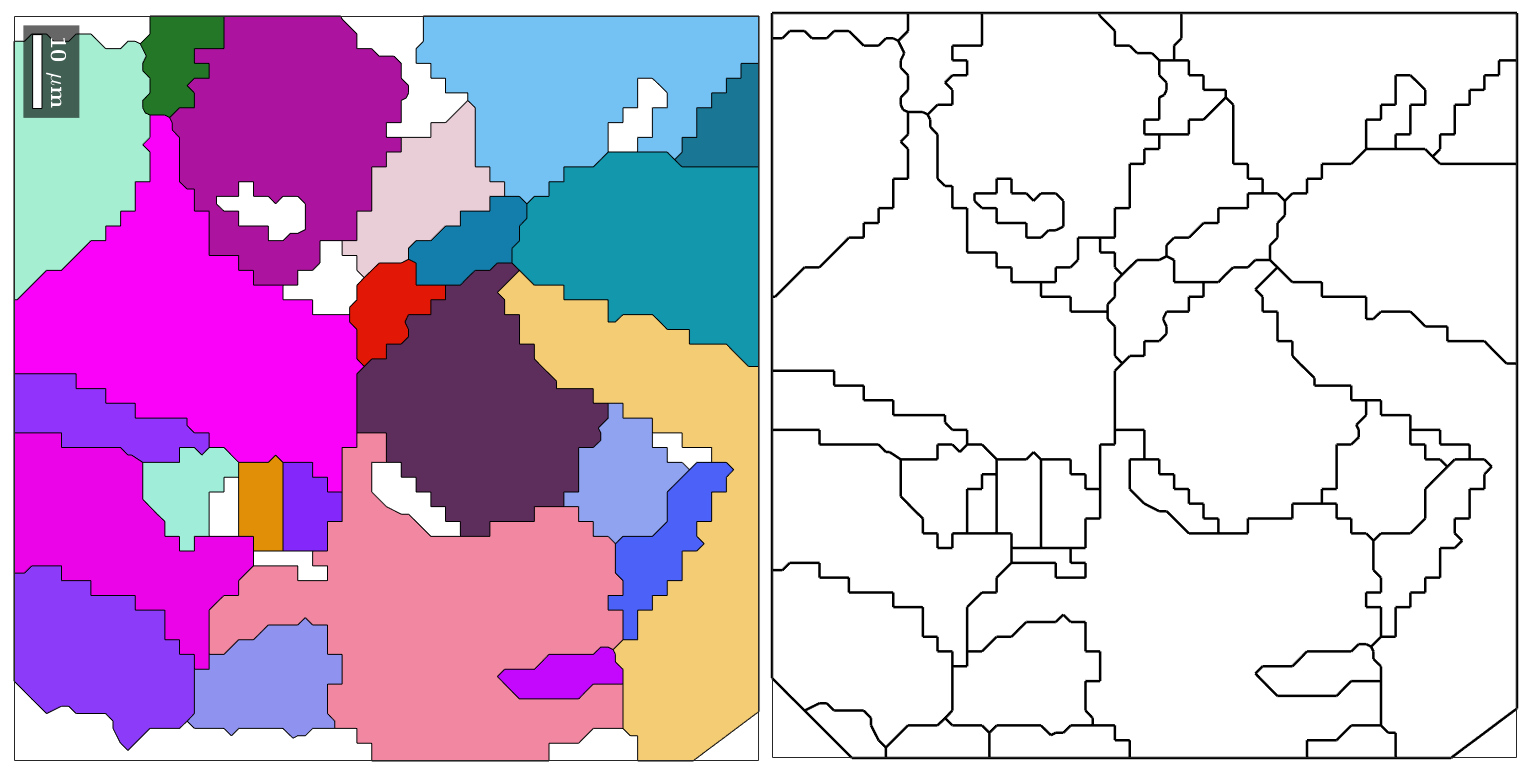}
        \caption{}
        \label{fig:pred_100ms}
    \end{subfigure}
    \hfill
    \begin{subfigure}{0.32\linewidth}
        \centering
        \includegraphics[width=\linewidth]{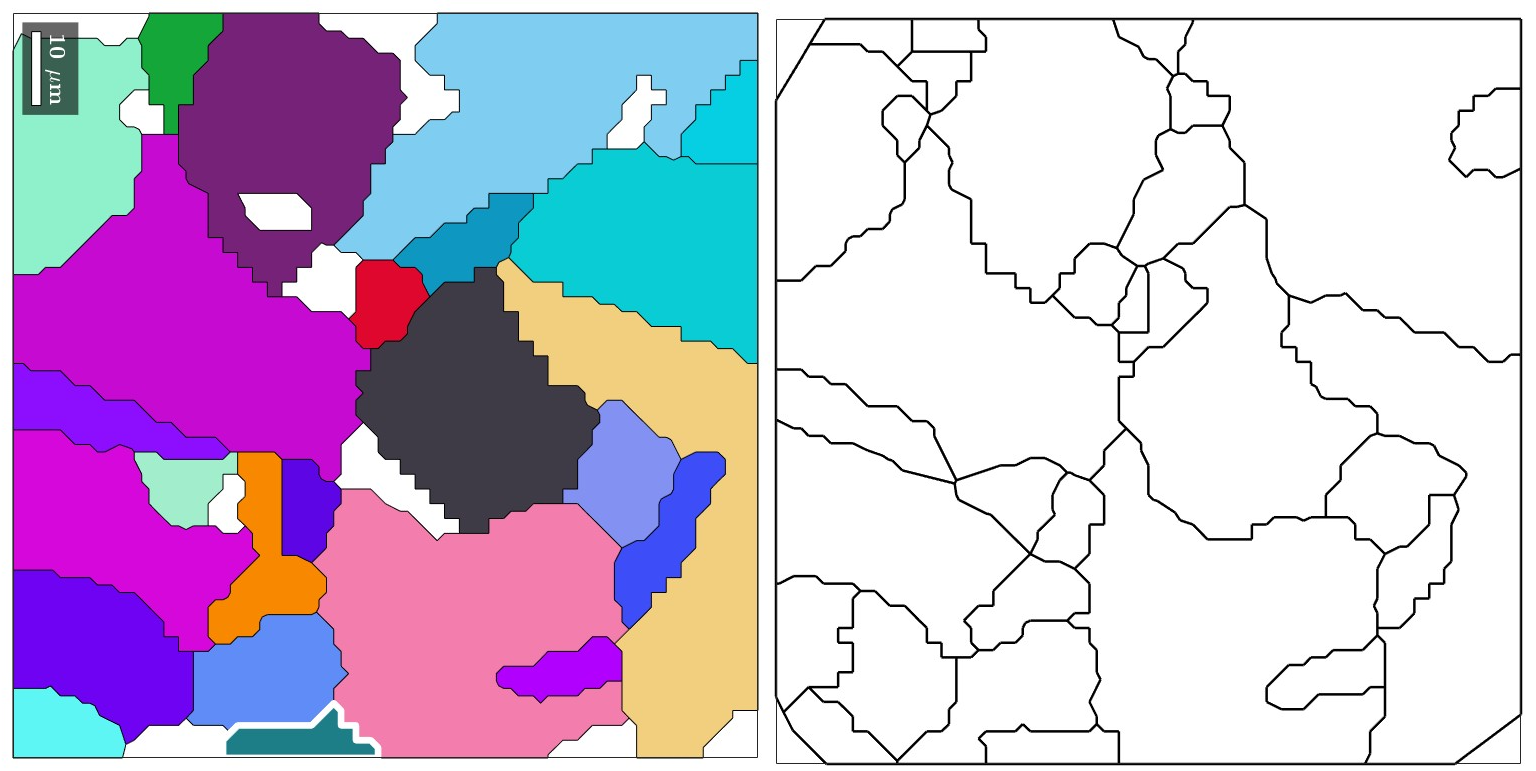}
        \caption{}
        \label{fig:pred_25ms}
    \end{subfigure}

    \vspace{0.3cm}

    \begin{subfigure}{0.32\linewidth}
        \centering
        \includegraphics[width=\linewidth]{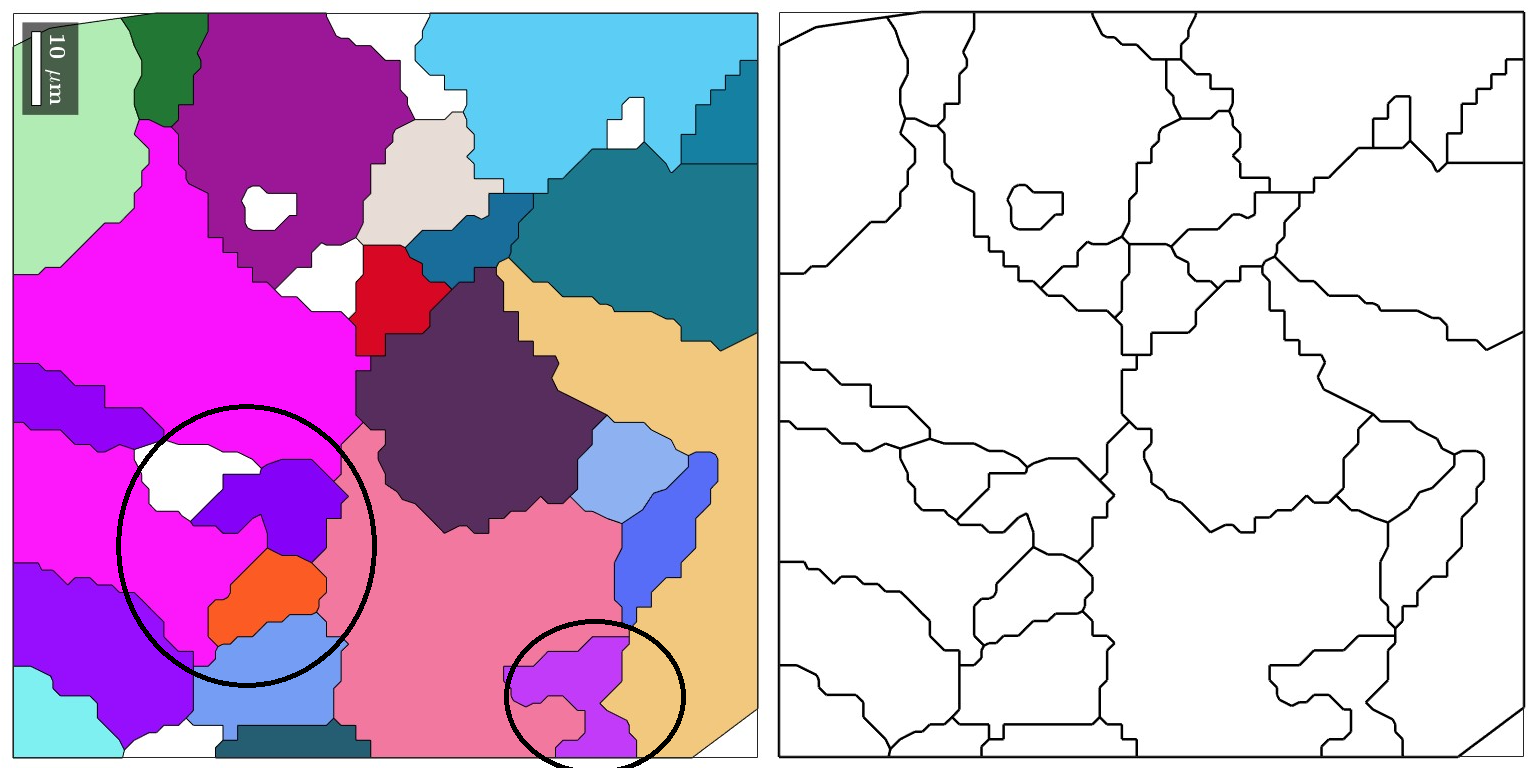}
        \caption{}
        \label{fig:pred_5ms}
    \end{subfigure}
    \hfill
    \begin{subfigure}{0.32\linewidth}
        \centering
        \includegraphics[width=\linewidth]{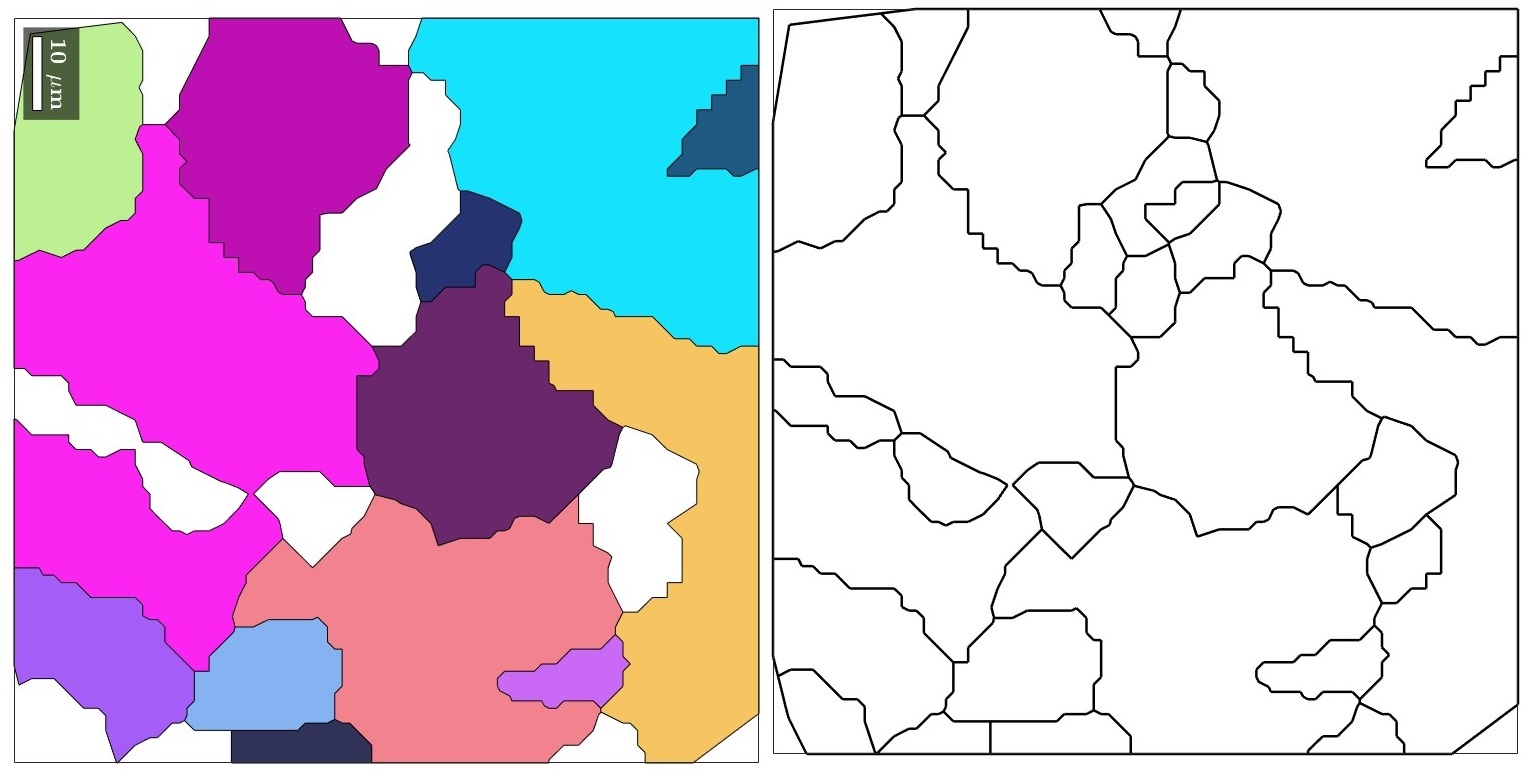}
        \caption{}
        \label{fig:pred_2.5ms}
    \end{subfigure}
    \hfill
    \begin{subfigure}{0.32\linewidth}
        \centering
        \includegraphics[width=\linewidth]{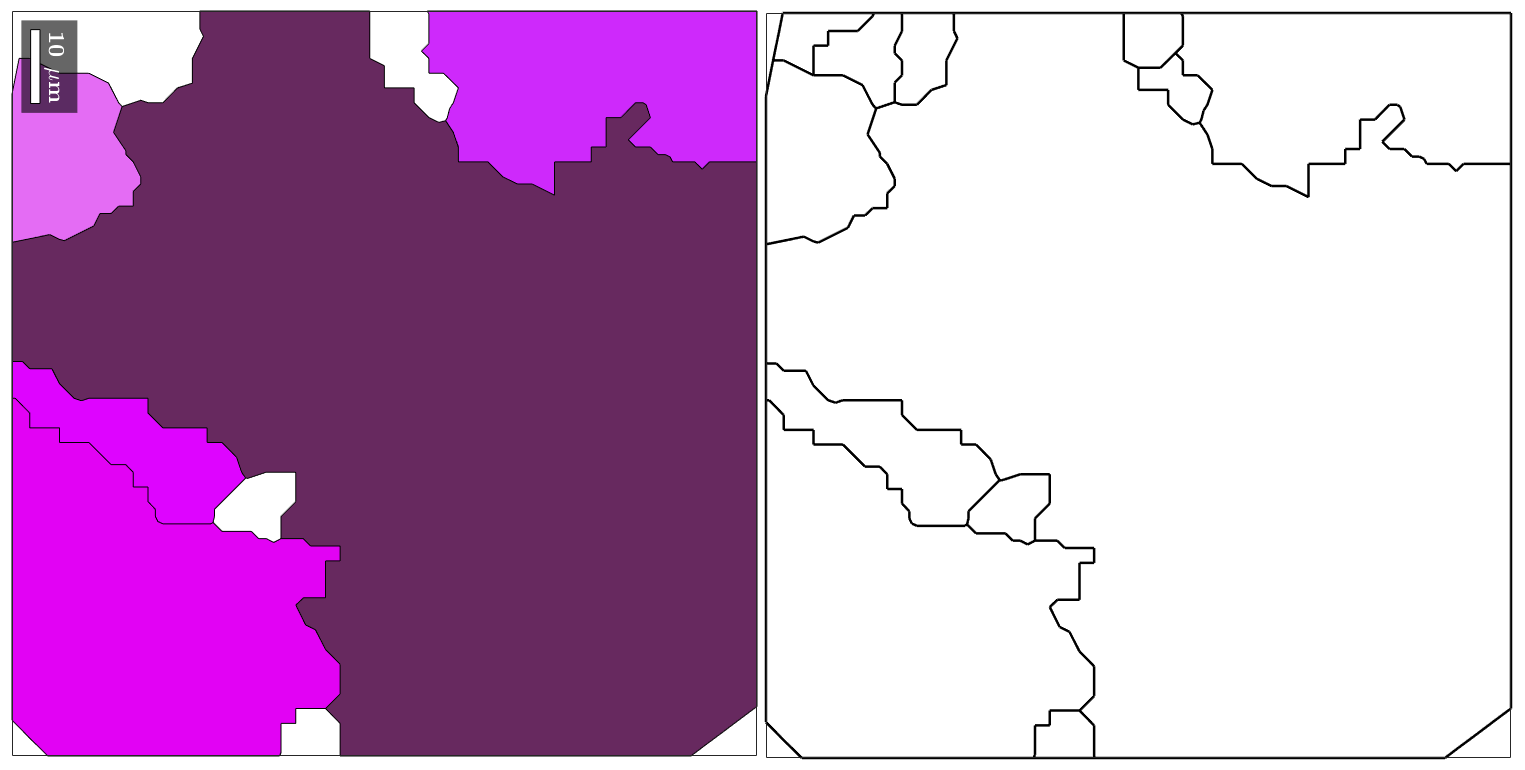}
        \caption{}
        \label{fig:pred_1ms}
    \end{subfigure}

    \caption{Comparison of true and predicted orientation maps. (a) Ground truth orientation map generated using MTEX with experimentally recorded Kikuchi patterns. (b) Predicted orientation map using CNN for 100 ms patterns. (c) Prediction on 25 ms pattern. (d) Prediction on 5 ms patterns. (e) Prediction on 2.5 ms patterns. (f) Prediction on 1 ms exposure time patterns.}
    \label{fig:orientation_maps}
\end{figure}

The CNN model was pre-trained on synthetic patterns generated from Kikuchipy library and then fine-tuned using the experimental patterns for varying scan speeds (i.e, on 100 ms, 25 ms, 5 ms, 2.5 ms and 1 ms exposure times per pattern). The predictions of CNN model for a given pattern are expressed by Euler angles, which correspond to the orientation of grain at the point which is color-coded to form a final orientation map for all the points on the specimen.
The results of CNN model are shown in the Figure \ref{fig:orientation_maps} for varying scan speeds as compared to the ground truth predictions. The ground truth predictions are determined by Hough transform method.

The orientation map shown in Figure \ref{fig:true_gb} was generated using MTEX. In accordance with the default settings, any misorientation within 15 degrees is treated as a single grain. 
Figures \ref{fig:pred_100ms} and \ref{fig:pred_25ms} show the orientations predicted for 100 ms and 25 ms exposure times, respectively. On comparing against the orientation map obtained by performing indexing for 100 ms patterns (Fig. \ref{fig:true_gb}), it is evident that the orientations of most grains have been accurately predicted with few exceptions, however, the predictions are not as good for the shorter exposure time. The predicted orientation map for patterns at 5 ms, 2.5 ms and 1 ms exposure time are shown in Figures \ref{fig:pred_5ms}, \ref{fig:pred_2.5ms} and \ref{fig:pred_1ms}, respectively.

\begin{figure}[H]
    \centering
    \includegraphics[width=0.5\linewidth]{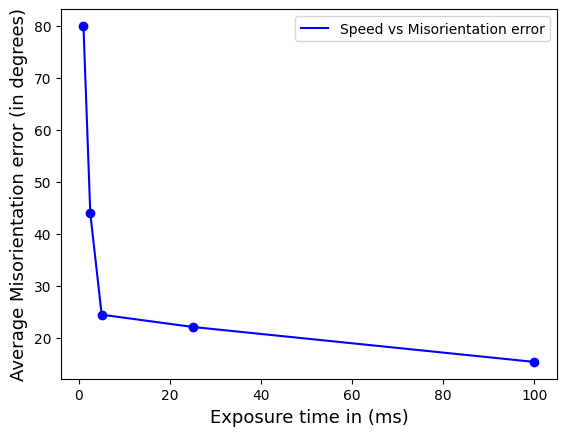}
    \caption{Predictive accuracy of CNN model (misorientation error vs exposure time). As the exposure time decreases, the misorientation error between true and predicted labels increases. It is evident that the CNN model performs well for the patterns up to 5 ms exposure time. But the predictive accuracy is significantly poor for the patterns obtained at short exposure times, e.g. 2.5 ms and 1 ms.}
    \label{fig:accuracy vs time}
\end{figure}

We measured accuracy in terms of average misorientation error between true and predicted Euler angles -- lower the misorientation error better the predictions. Typically, orientations within the range of 10-15 degrees are considered as the orientations of the same grain. Our CNN model showed reasonable accuracy for indexing the high-quality patterns obtained and 100 ms exposure time, the average misorientation error between true and predicted orientation by model is about 10-12 degrees, which is satisfactory. However, the accuracy can be further improved by tuning hyperparameters, using a better loss function and architecture, and training it on a more diverse dataset. But the initial objective of our case study of CNN model is to identity the amount of noise till which the model can still map the pattern to their corresponding orientation. Therefore, we fine-tuned the model for different scan speeds to identify the speed vs accuracy trade-off.
It is clear from the above results in the above Figure \ref{fig:orientation_maps} that the CNN model struggles to find features to map orientation of patterns at very short exposure times i.e. 2.5 ms and 1 ms. There is a significant drop in the accuracy of orientation indexing for patterns obtained at 2.5 ms and 1 ms exposure times as shown in the Figure \ref{fig:accuracy vs time}. 

\noindent Since, as evident from the aforementioned, the CNN model can be used to do the orientation indexing of the noisy Kikuchi patterns, but it may still struggle if the noise level in the patterns increases beyond a limit. Hence to push the limits of orientation indexing beyond that limitation of CNN-model we decided to leverage the power of the generative diffusion model to restore the patterns obtained at short exposure times i.e, 1 ms and 0.5 ms exposure times in our case. Once the quality of the patterns improves after post-processing by the model, the patterns can be further processed by the Hough indexing method to obtain the orientations. We would like to emphasise that orientation indexing at shorter exposure times would lead to much shorter total times of data capture of EBSD. Therefore, there is a strategic interest in being able to predict the correct orientations for short exposure times.

\section{Generative diffusion modelling for Kikuchi pattern quality enhancement}
\begin{figure}[H]
        \centering
        \includegraphics[trim={0 6cm 0 6cm},clip,width=0.9\linewidth]{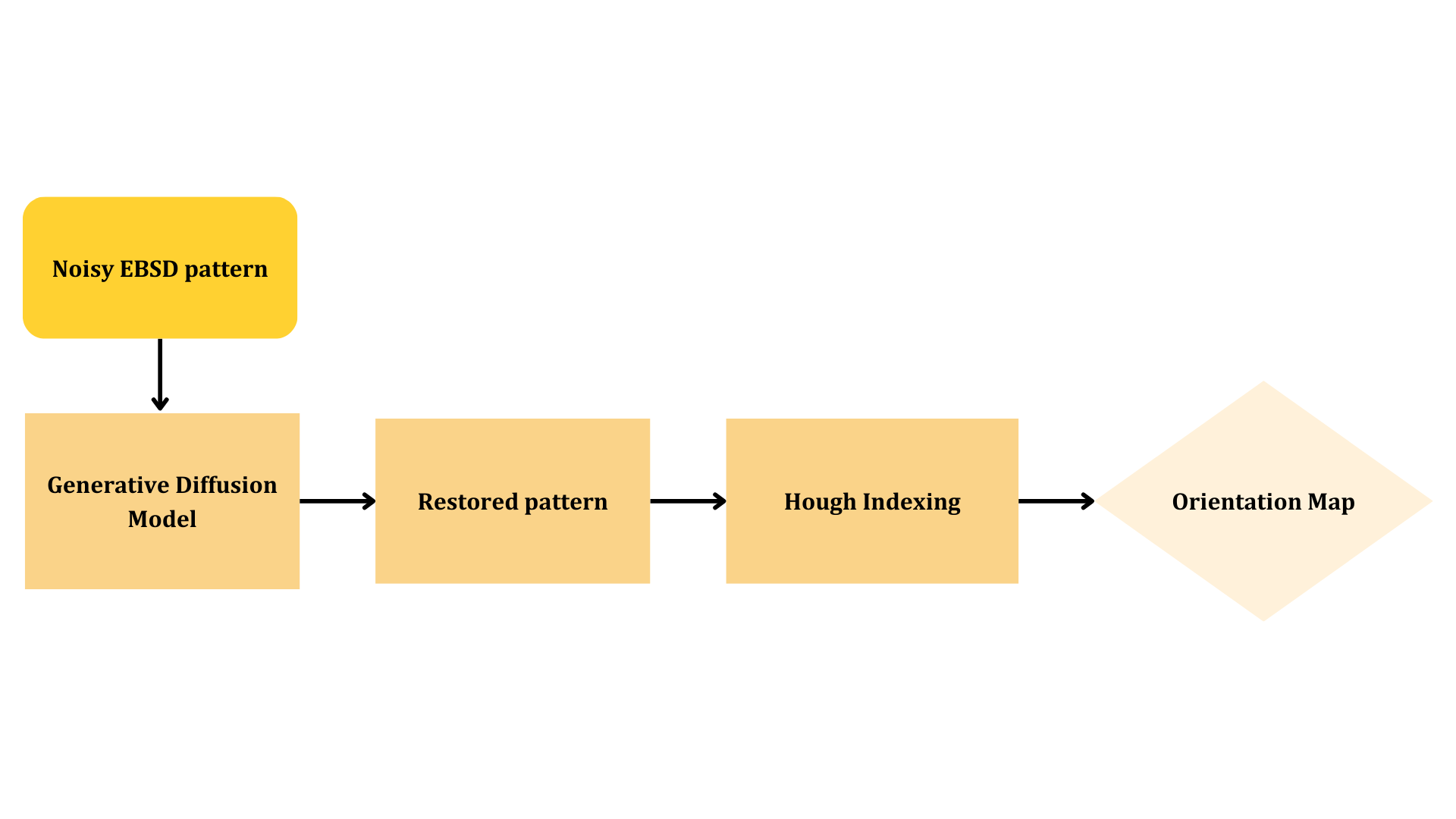}
        \caption{The generative diffusion model-based pipeline for enhanced pattern indexing.}
        \label{Diffusion flow chart}
\end{figure}

\noindent The objective of generative diffusion model is to enhance the quality of short exposure time patterns such that the patterns can further be analyzed using Hough transform based method to provide improved and reliable orientation indexing results. The flow chart of diffusion based enhanced pattern indexing method is shown in the Figure \ref{Diffusion flow chart}.

\subsection{Methodology}

\noindent In the current work, we explored the capabilities of generative diffusion models to enhance the quality of high-speed scan EBSD patterns by training the models to generate the high-quality patterns by denoising starting from the pure Gaussian noise guided by conditioning on short exposure time patterns.

\noindent In recent years, many supervised and zero-shot variants of diffusion models have been developed for tasks such as image super-resolution and image restoration \cite{saharia2022image, saharia2022palette, rombach2022high}. All these models follow the principle of conditional image generation \cite{batzolis2021conditional}, which involves generating a high-quality image by conditioning the diffusion model on a low-quality image.

Diffusion models are trained through two complementary processes - the \textit{forward diffusion} and the \textit{reverse diffusion}. In the forward diffusion process, high-quality training images are progressively corrupted by adding Gaussian noise over a series of time steps. This transforms the original data distribution into a standard Gaussian distribution. To do this, the noise is added to a clean image $\mathbf{x_0}$ over $T$ time steps according to a predefined variance schedule $\{\beta_t\}_{t=1}^T$ i.e., linearly increasing in our case. At each time step $t$, the noised image $\mathbf{x_t}$ is generated from the previous image $\mathbf{x_{t-1}}$ using a Gaussian distribution as \cite{ho2020denoising}

\begin{equation}
    q(\mathbf{x}_t \mid \mathbf{x}_{t-1}) = \mathcal{N}(\mathbf{x}_t; \sqrt{1 - \beta_t} \, \mathbf{x}_{t-1}, \beta_t \mathbf{I}),
    \label{Eq:forward_diffusion}
\end{equation}

where, $t$ is a random time step sampled uniformly from $[1, T]$, $\mathbf{x}_t$ is the noisy version of the data at time step $t$ and $\beta_t$ is the constant obtained from the noise schedule. During training, the model learns the reverse diffusion process i.e. to iteratively denoise a sample starting from pure Gaussian noise. At each time step, the model is trained to remove only a small portion of the noise, thereby gradually reconstructing the clean image. In conditional diffusion models, the denoising at each step is guided not only by the noisy sample from the previous timestep but also by an external condition which is the low-quality raw Kikuchi pattern in our case. This conditional input ensures that the generation is informed by relevant structural information to be restored.

The reverse diffusion process is modeled as a Markov chain with Gaussian transitions. At each time step $t$, the model predicts the mean $\mu_\theta(\mathbf{x}_t, t, \mathbf{y})$ conditioned on the low quality input $\mathbf{y}$, while the variance $\tilde{\beta}_t$ is kept fixed and derived from the forward noise schedule. Hence, the reverse process is defined as \cite{ho2020denoising}

\begin{equation}
    p_\theta \left( \mathbf{x}_{t-1} \mid \mathbf{x}_t, \mathbf{y} \right) = \mathcal{N}\left( \mathbf{x}_{t-1}; \mu_\theta \left( \mathbf{x}_t, t, \mathbf{y} \right), \tilde{\beta}_t \mathbf{I} \right).
\end{equation}

Although the reverse diffusion process is formally expressed as a Gaussian distribution with a predicted mean $\mu_\theta(\mathbf{x}_t, t, \mathbf{y})$, in practice, the model is trained to predict the noise $\epsilon_\theta(\mathbf{x}_t, t, \mathbf{y})$ that was added to the clean image $\mathbf{x}_0$ during the forward process. This approach simplifies training and improves stability, as demonstrated in the original DDPM framework \cite{ho2020denoising}.

The mean squared error between the true added noise at a step and the noise predicted by the diffusion model is considered to be the loss at each time step as given by \cite{ho2020denoising}

\begin{equation}
\mathcal{L}_{\text{Diffusion}}(\mathbf{\theta}) = {E}_{t, \mathbf{x}_0, \epsilon} \left[ \left\| \epsilon - \epsilon_\mathbf{\theta} \left(\mathbf{x}_t, \mathbf{y}, t \right) \right\|^2 \right],
\end{equation}
where, $\textbf{x}_0$ is the original data sample (e.g., the clean image), $\epsilon$ is the true noise added at time step $t$, $\epsilon_\mathbf{\theta} \left(\mathbf{x}_t, \mathbf{y}, t \right)$ is the noise predicted by the model parameterized by $\mathbf{\theta}$ and $\|\cdot\|$ represents the Euclidean norm.

\begin{figure}[H]
        \centering
        \includegraphics[width=0.6\linewidth]{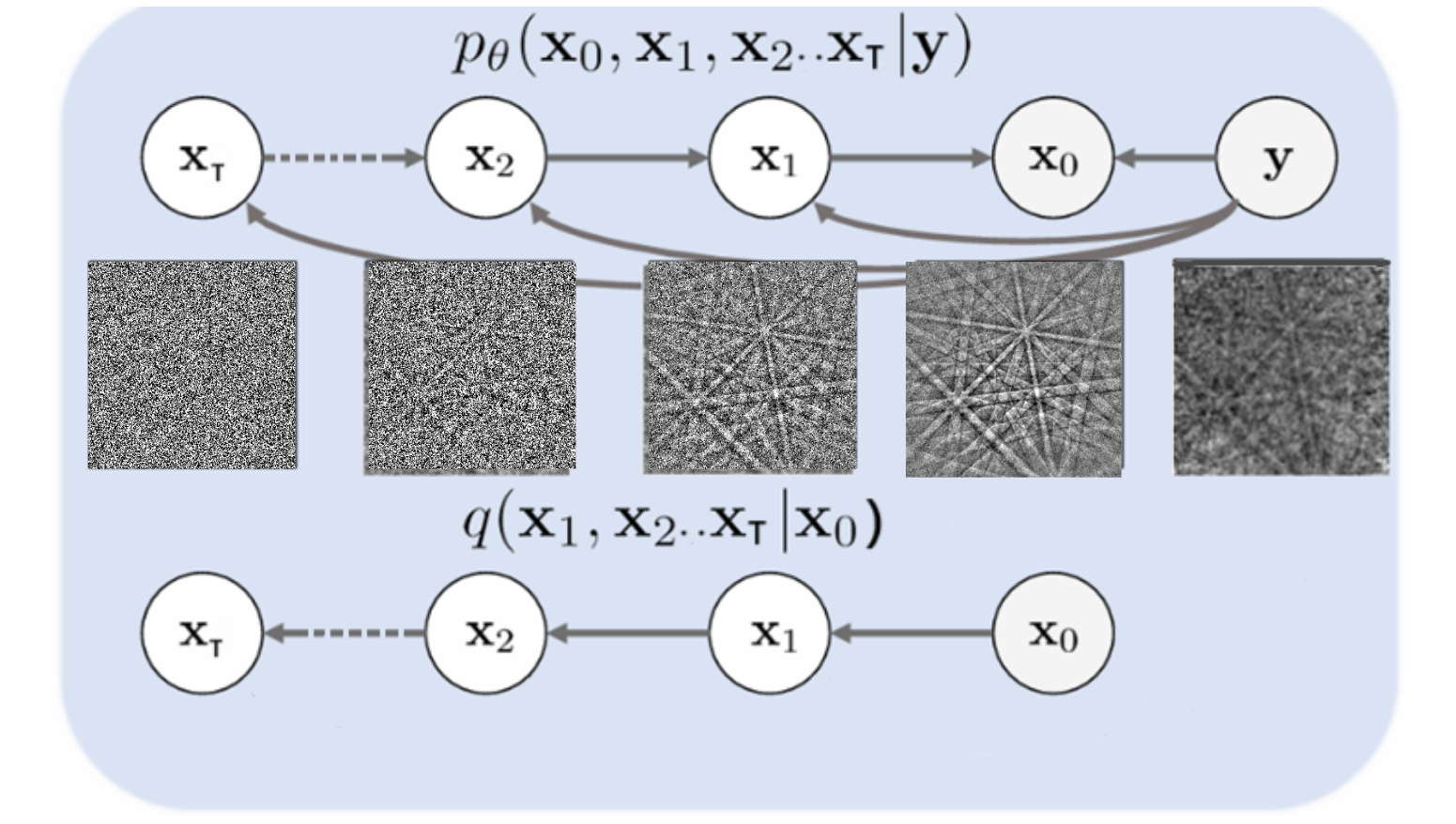}
        \caption{Forward and reverse diffusion with conditioning, after Kawar et al. \cite{kawar2022denoising}.}
        \label{fig:Forward and reverse diffusion}
\end{figure}

An overview of the complete process is illustrated in Figure \ref{fig:Forward and reverse diffusion}, where $q$ denotes the forward diffusion, $p_\mathbf{\theta}$ represents the reverse process, and $y$ is the input conditioning pattern. 

\begin{figure}[H]
    \centering
    \includegraphics[width=0.8\linewidth]{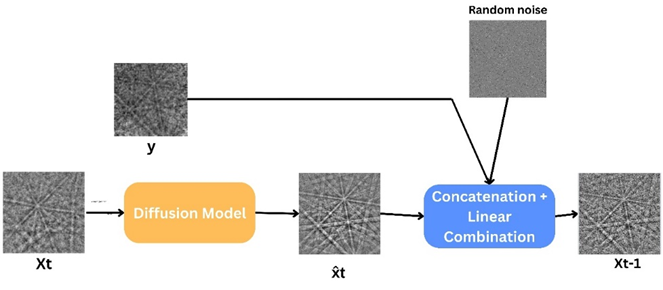}
    \caption{Reverse diffusion applied at time step $t$ for generation of noise-free Kikuchi pattern $\hat{\mathbf{x}}$ given the previous time-step image $\mathbf{x}_{t}$ and input low quality conditional pattern $\mathbf{y}$. Also, some small amount of noise is re-added to the model's output $\hat{\mathbf{x}}$ to obtain $\mathbf{x}_{t-1}$ during both training and sampling, except at the final stage of generation i.e, to obtain $\mathbf{x}_0$.}
    \label{fig:Reverse Diffusion}
\end{figure}

For conditional image generation, several conditioning strategies have been proposed in the literature, including concatenation--based conditioning \cite{saharia2022image}, cross--attention mechanisms \cite{lin2022cat}, and modulation techniques \cite{perez2018film}, depending on the model architecture and application context. In our approach, we adopted the fundamental and highly effective image conditioning strategy proposed in the work by Saharia et al. \cite{saharia2022image}. This method involves channel-wise concatenation of the input conditional image at the current time step with the noisy image from the previous step. The concatenated inputs are then fed into the diffusion model, which predicts the noise added in the image at the current step. In the present study, we have used the grayscale images of short exposure time noisy Kikuchi pattern as the input condition. The details of conditioning at an intermediate step of reverse diffusion is shown in the Figure \ref{fig:Reverse Diffusion}.

Once trained, the diffusion model is capable of refining EBSD patterns by iteratively denoising images. Since, the model learns to predict the noise added at each diffusion step during training, during inference, it serves as a conditional denoiser. Starting from pure Gaussian noise, it progressively removes noise to generate a high-quality Kikuchi pattern conditioned on a given low-quality input pattern. Conditional input ensures that the output is not a random sample from the data distribution but a restored, high-quality version of the degraded input.

Sampling for high-quality Kikuchi pattern during the reverse diffusion process is performed as follows. Let $\mathbf{x}_T \sim \mathcal{N}(0, I)$ be a sample from the standard Gaussian prior, and let $\epsilon_\theta(\mathbf{x}_t, \mathbf{y}, t)$ denote the model’s estimate of the noise component at time step $t$ parameterized by $\theta$, where $\mathbf{x}_t$ is the denoised image of pattern at time step $t$ and $\mathbf{y}$ is the input conditional low quality Kikuchi pattern. The reverse diffusion dynamics, as formulated in the DDPM framework \cite{ho2020denoising}, define a Markov chain as

\begin{equation}
\mathbf{x}_{t-1} = \frac{1}{\sqrt{\alpha_t}} \left( \mathbf{x}_t - \frac{1 - \alpha_t}{\sqrt{1 - \bar{\alpha}_t}} \epsilon_\theta(\mathbf{x}_t,\mathbf{y}, t) \right) + \sigma_t z, \quad z \sim \mathcal{N}(0, I),
\label{eq:ddpm}
\end{equation}

where, $\mathbf{x}_{t-1}$ is the denoised image at step $t-1$, $\alpha_t$, $\bar{\alpha}_t$, and $\sigma_t$ are parameters derived from a predefined variance (noise) schedule. The denoising process proceeds from $\mathbf{x}_T$ down to $\mathbf{x}_0$, guided at each timestep by the conditional low quality input pattern $\mathbf{y}$, which ensures that the output is structurally consistent with the degraded Kikuchi pattern provided.

\subsection{Architecture details}
\begin{figure}[H]
    \centering
    \includegraphics[width=0.8\linewidth]{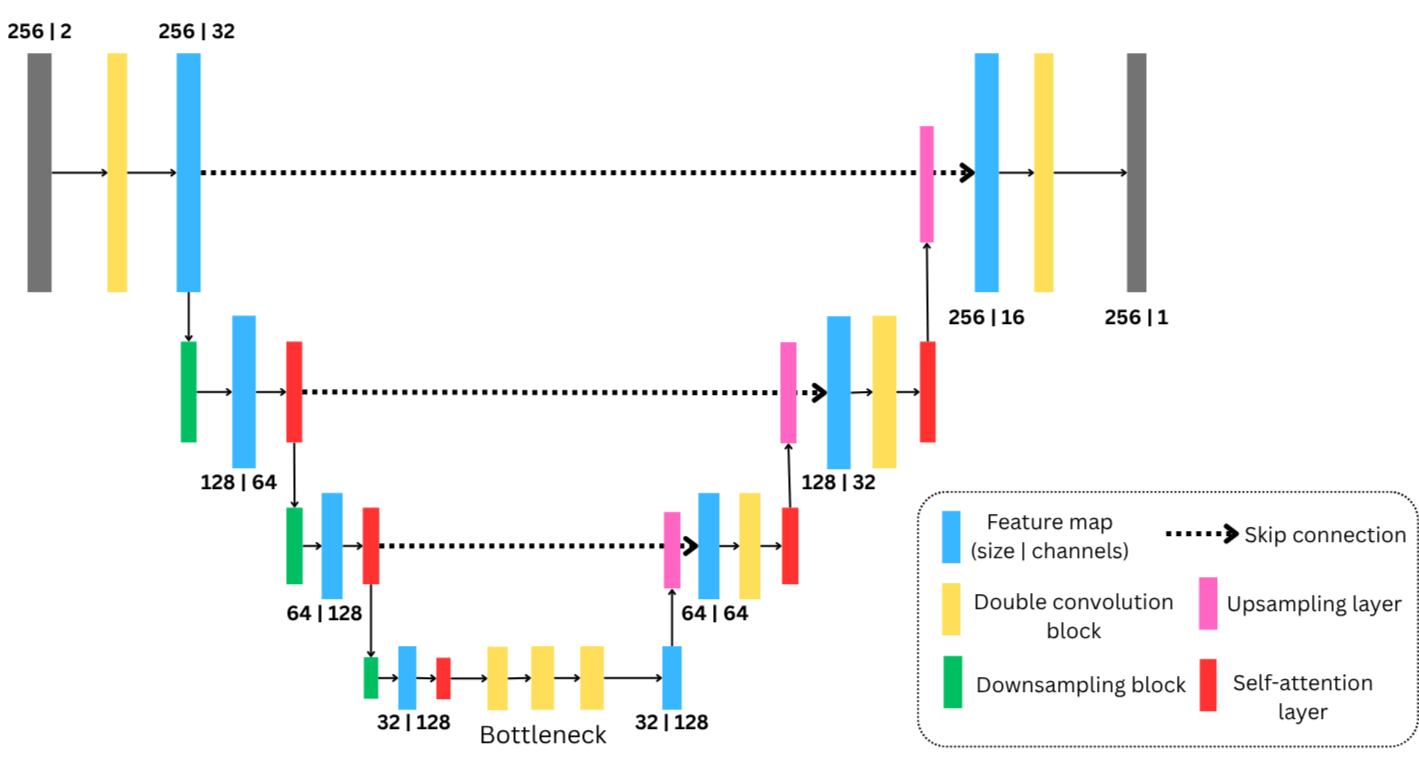}
    \caption{U-Net architecture diagram of our diffusion model.} 
    \label{fig:Unet diagram}
\end{figure}

\noindent Our architecture of the U-Net model \cite{ronneberger2015u} for the diffusion framework is inspired by the architecture proposed in the original work wherein DDPM was proposed \cite{ho2020denoising}. This architecture closely follows the backbone of PixelCNN++ \cite{salimans2017pixelcnn++}, which is a U-Net-based architecture derived from a wide ResNet \cite{he2016deep}. While the original architecture was designed to handle RGB images, we modified the U-Net architecture to handle grayscale images of Kikuchi patterns of size 256$\times$256 and reduced some layers. The model follows a multi-scale encoder-decoder structure with self-attention layers and time-dependent conditioning using sinusoidal positional embeddings. The detailed architecture diagram of U-Net is shown in the Figure \ref{fig:Unet diagram}. The encoder extracts hierarchical features while progressively reducing the resolution from 256 $\times$ 256 to 128 $\times$ 128, 64 $\times$ 64, and finally 32 $\times$ 32 via a down sampling block which consists of double convolution layers and a down sampling layer. The block of double convolution layers maintains the resolutions but increases the number of output channels. Self-attention layers are applied after each down sampling block at 128 $\times$ 128, 64 $\times$ 64, and 32 $\times$ 32 resolution levels to capture long-range dependencies within the image of Kikuchi patterns. A three-layer bottleneck of the blocks of double convolution layers encodes the most abstract representation of the input. The decoder follows the mirror architecture of encoder and up-samples the features back to 256 $\times$ 256 while maintaining spatial details through skip connections.

The model receives two-channel inputs (noisy image from the previous step + conditioning image of the pattern) as part of our conditioning strategy and reconstructs a single-channel restored output. Group normalization has been used for regularization and better stability during training. 2D dropout between the blocks of double convolution layers have also been experimented. GELU activation function has been used, which provides smoother gradients than ReLU during training \cite{hendrycks2016gaussian}. To integrate the diffusion time step $t$, a sinusoidal positional encoding is injected into every up sampling and down sampling block, ensuring the model adapts dynamically to different noise levels. The proposed architecture of the model has the total of 5,881,569 trainable parameters.

\section{Training and hyper-parameters}
\begin{figure}[H]
    \centering
    \begin{subfigure}{0.45\linewidth} 
        \centering
        \includegraphics[width=\linewidth]{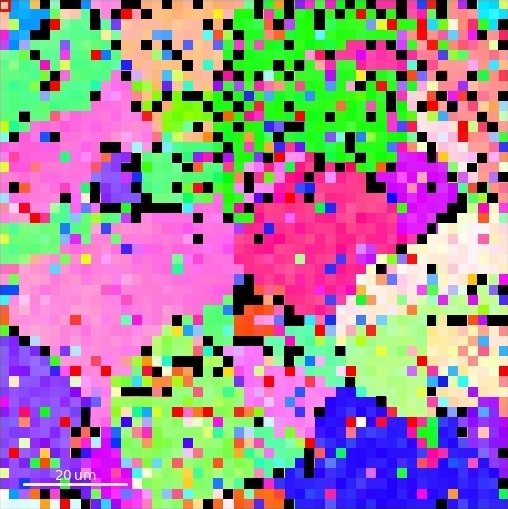}
        \caption{}
        \label{fig:sampled_points}
    \end{subfigure}
    \hfill
    \begin{subfigure}{0.45\linewidth}
        \centering
        \includegraphics[width=\linewidth]{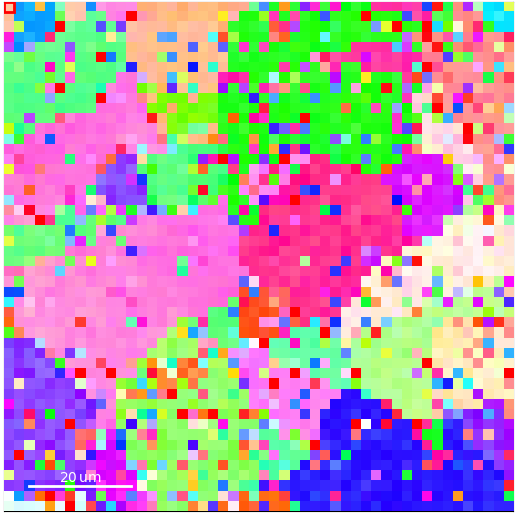}
        \caption{}
        \label{fig:IPF 1ms_2}
    \end{subfigure}
    
    \caption{(a) Sampled points for training are shown by black pixels on the orientation map of experimental(raw) 1 ms patterns. (b) Orientation map of the raw 1 ms patterns for the comparison.}
    \label{fig:IPF_comparison}
\end{figure}

\noindent The proposed diffusion model was trained on an NVIDIA A100 Tensor Core GPU using paired EBSD patterns i.e. high-quality reference patterns acquired at a 100 ms exposure time and corresponding degraded patterns captured at a 1 ms exposure time. To construct the training and validation datasets, only 10 $\%$ of the total points from a 50 $\times$ 50 spatial grid were sampled. The remaining 90 $\%$ of the data points were withheld and used exclusively for testing to evaluate the model’s generalization and restoration performance on the entire scan map. Data augmentation techniques were applied to the sampled training points to improve generalization and enrich the training data despite the small sample size. The sampled training points are indicated as black pixels in Figure~\ref{fig:sampled_points}. Points apart from black pixels were not seen by the model throughout the training. The corresponding low-quality orientation map obtained at 1 ms exposure can be seen in Figure~\ref{fig:IPF 1ms_2} for comparison. During training the samples were visualized after each 10 epochs to inspect for the over-fitting and mode coverage. The training spanned 150 epochs and took 11-12 hours overall, in which the checkpoints of model weights were saved every 10 epochs. The AdamW optimizer \cite{loshchilov2017decoupled} with the default value of weight decay (i.e, 0.01) was used. A lower learning rate of $2 \times 10^{-4}$ was used to avoid instability during training. An exponential moving average (EMA) with a decay factor of 0.995 was applied to the model parameters to improve stability. The conditioning strategy allows the model to predict noise at each step and refine the image progressively to generate a higher quality pattern which matches the conditional image. The loss function used is the mean squared error between the true added noise and the predicted noise at each step. A linear noise variance schedule, starting from $\beta_1 = 10^{-4}$ to $\beta_T = 0.02$ over $T = 500$ time steps was applied to noise the images as given by Eq. \ref{Eq:forward_diffusion}.

\section{Results}

\subsection{Restored pattern quality}

\begin{figure}[H]
    \centering
    \begin{subfigure}{0.18\linewidth}
        \centering
        \includegraphics[width=\linewidth]{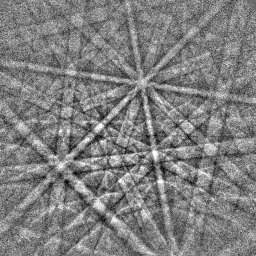}
    \end{subfigure}
    \begin{subfigure}{0.18\linewidth}
        \centering
        \includegraphics[width=\linewidth]{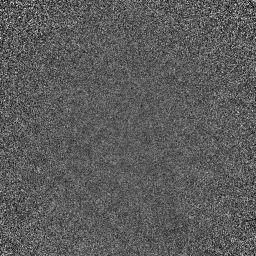}
    \end{subfigure}
    \begin{subfigure}{0.18\linewidth}
        \centering
        \includegraphics[width=\linewidth]{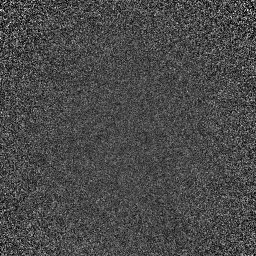}
    \end{subfigure}
    \begin{subfigure}{0.18\linewidth}
        \centering
        \includegraphics[width=\linewidth]{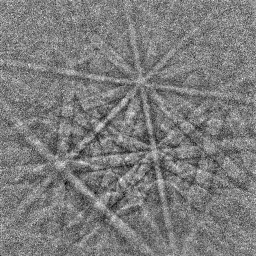}
    \end{subfigure}
        \begin{subfigure}{0.18\linewidth}
        \centering
        \includegraphics[width=\linewidth]{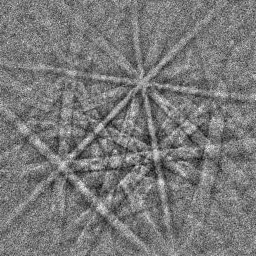}
    \end{subfigure}

    \begin{subfigure}{0.18\linewidth}
        \centering
        \includegraphics[width=\linewidth]{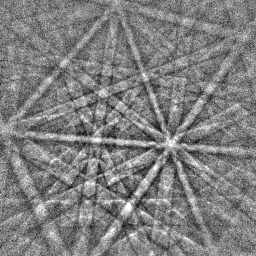}
    \end{subfigure}
    \begin{subfigure}{0.18\linewidth}
        \centering
        \includegraphics[width=\linewidth]{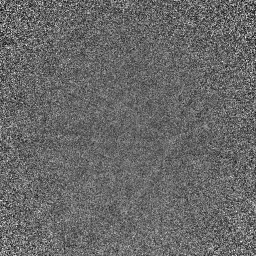}
    \end{subfigure}
    \begin{subfigure}{0.18\linewidth}
        \centering
        \includegraphics[width=\linewidth]{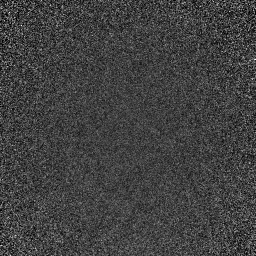}
    \end{subfigure}
    \begin{subfigure}{0.18\linewidth}
        \centering
        \includegraphics[width=\linewidth]{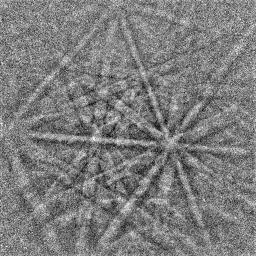}
    \end{subfigure}
        \begin{subfigure}{0.18\linewidth}
        \centering
        \includegraphics[width=\linewidth]{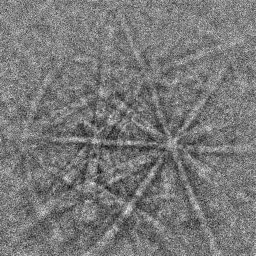}
    \end{subfigure}

    \begin{subfigure}{0.18\linewidth}
        \centering
        \includegraphics[width=\linewidth]{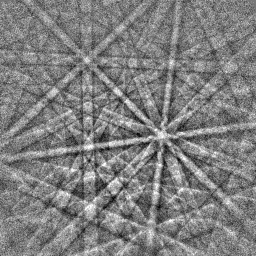}
    \end{subfigure}
    \begin{subfigure}{0.18\linewidth}
        \centering
        \includegraphics[width=\linewidth]{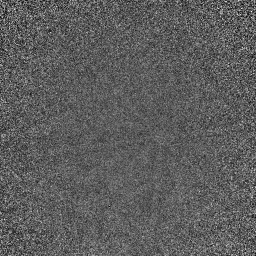}
    \end{subfigure}
    \begin{subfigure}{0.18\linewidth}
        \centering
        \includegraphics[width=\linewidth]{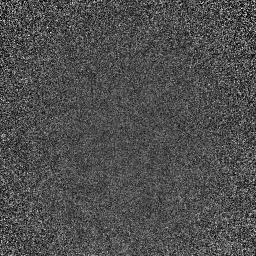}
    \end{subfigure}
    \begin{subfigure}{0.18\linewidth}
        \centering
        \includegraphics[width=\linewidth]{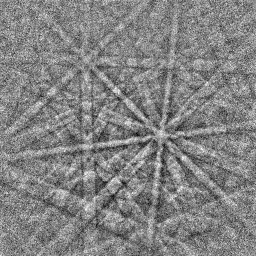}
    \end{subfigure}
        \begin{subfigure}{0.18\linewidth}
        \centering
        \includegraphics[width=\linewidth]{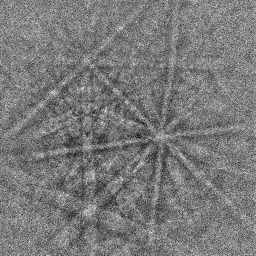}
    \end{subfigure}

    \begin{subfigure}{0.18\linewidth}
        \centering
        \includegraphics[width=\linewidth]{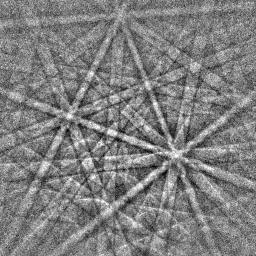}
        \caption{}
    \end{subfigure}
    \begin{subfigure}{0.18\linewidth}
        \centering  
        \includegraphics[width=\linewidth]{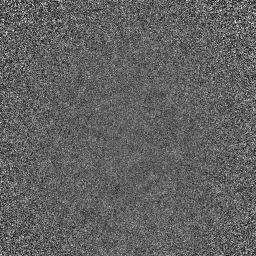}
        \caption{}
    \end{subfigure}
    \begin{subfigure}{0.18\linewidth}
        \centering
        \includegraphics[width=\linewidth]{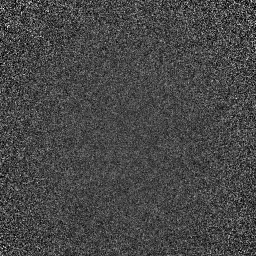}
        \caption{}
    \end{subfigure}
    \begin{subfigure}{0.18\linewidth}
        \centering
        \includegraphics[width=\linewidth]{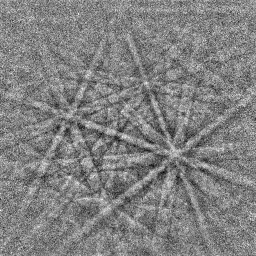}
        \caption{}
    \end{subfigure}
    \begin{subfigure}{0.18\linewidth}
        \centering
        \includegraphics[width=\linewidth]{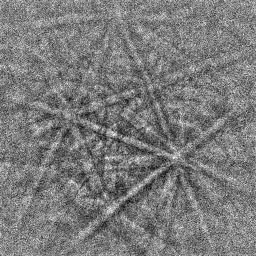}
        \caption{}
    \end{subfigure}

    \caption{1 ms exposure time patterns enhanced by the diffusion model. (a) The Kikuchi patterns captured at 100 ms using experiment (ground truth). (b)  Kikuchi patterns experimentally captured at 1 ms exposure time (or raw 1 ms patterns). These patterns are used as input to the diffusion model for restoring the Kikuchi bands. (c) Kikuchi patterns experimentally captured at 0.5 ms exposure time (or raw 0.5 ms patterns). (d) 1 ms patterns restored by the diffusion model. (e) 0.5 ms patterns restored by the diffusion model.}
    
    \label{fig:diffusion_collage}
\end{figure}

\noindent The results of the diffusion model generating high-quality restored Kikuchi patterns by conditioning on short exposure time (1 ms) patterns are shown in Figure \ref{fig:diffusion_collage}. In the raw, unprocessed Kikuchi patterns of 1 ms exposure time, many important features are typically absent due to the low signal-to-noise ratio. However, our model effectively restores much of this lost information, making the previously faint or absent features, such as key Kikuchi bands and poles, clearly visible after post-processing.

\noindent The raw patterns captured for 1 ms exposure time were processed through our generative diffusion model which is trained to restore the missing features of Kikuchi patterns. These generated patterns are used to create the orientation maps by the Hough transform-based orientation indexing method in Python using the Kikuchipy library \cite{hakon_wiik_anes_2024_11432173}.

\begin{figure}[H]
    \centering
    \begin{subfigure}{0.3\linewidth}
        \centering
        \includegraphics[width=\linewidth]{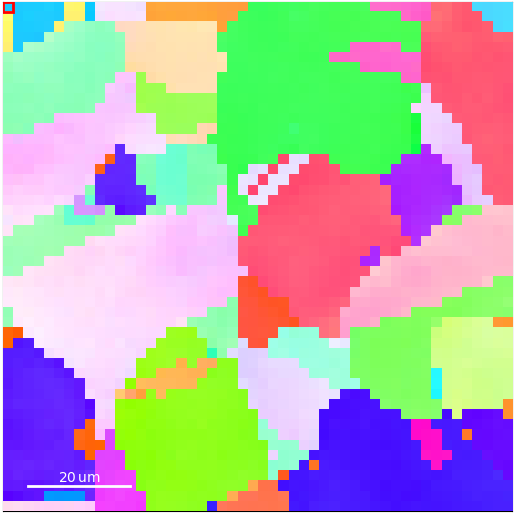}
        \caption{ }
        \label{fig:DDPM_IPF_100ms 2}
    \end{subfigure}
    \hfill
    \begin{subfigure}{0.3\linewidth}
        \centering
        \includegraphics[width=\linewidth]{1ms_Patterns_HI_orientation_map.png}
        \caption{}
        \label{fig:DDPM_IPF 1ms}
    \end{subfigure}
    \hfill
    \begin{subfigure}{0.3\linewidth}
        \centering
        \includegraphics[width=\linewidth]{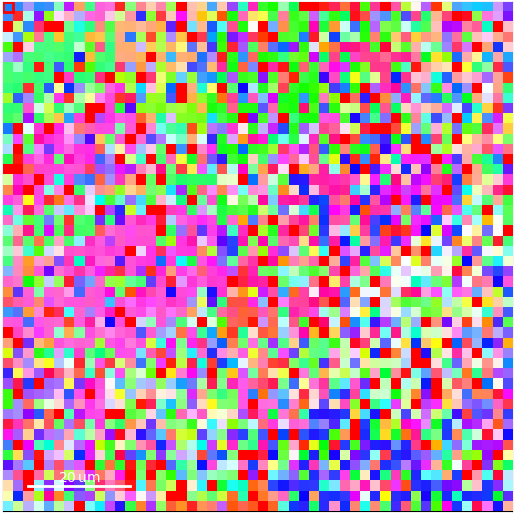}
        \caption{}
        \label{fig:DDPM_raw_0.5ms 1}
    \end{subfigure}

    \vspace{1em} 

    \begin{subfigure}{0.3\linewidth}
        \centering
        \includegraphics[width=\linewidth]{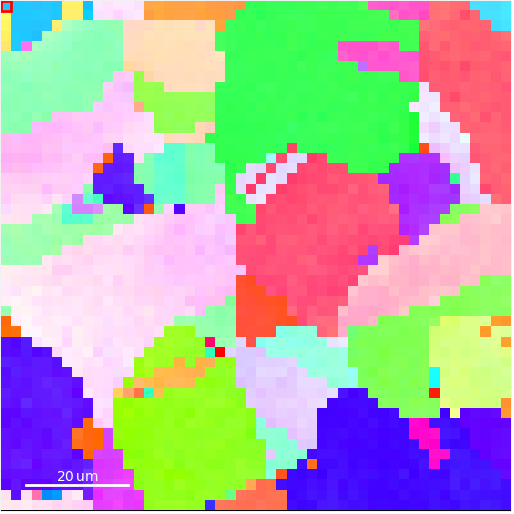}
        \caption{}
        \label{fig:DDPM_restored IPF}
    \end{subfigure}
    \hspace{1em}
    \begin{subfigure}{0.3\linewidth}
        \centering
        \includegraphics[width=\linewidth]{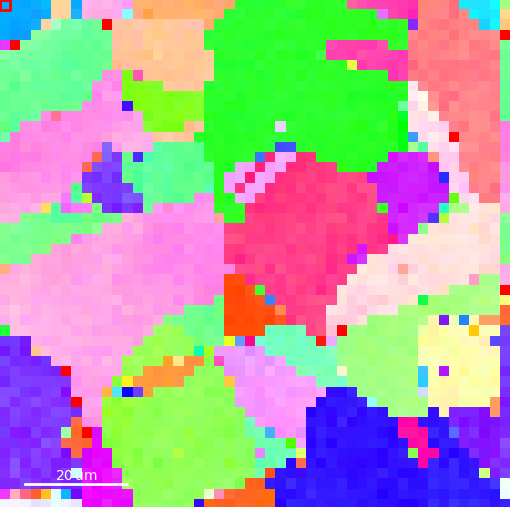}
        \caption{}
        \label{fig:DDPM_restored_0.5ms 1}
    \end{subfigure}

    \caption{Orientation map obtained by Hough indexing on raw experimental patterns for (a) 100 ms exposure time, (b) 1 ms exposure time, (c) 0.5 ms exposure time. Orientation map based on restored Kikuchi patterns for (d) 1 ms exposure time (e) 0.5 ms exposure time. Significant improvement is achieved in the orientation indexing of Kikuchi patterns as compared to the orientation map of raw 1 ms and 0.5 ms exposure time patterns after passing them for processing through our trained conditional generative diffusion model.}
    \label{fig:IPF_comparison}
\end{figure}

The orientation map obtained using Hough transform-based orientation indexing on the high-quality patterns captured for the 100 ms exposure time as shown in Figure \ref{fig:DDPM_IPF_100ms 2} serves as the ground truth reference for comparison. The raw orientation map corresponding to the patterns captured for 1 ms exposure time in Figure \ref{fig:DDPM_IPF 1ms} shows considerable wrong indexing and orientation noise due to poor pattern quality. Similarly, the patterns captured for 0.5 ms exposure shown in Figure \ref{fig:DDPM_raw_0.5ms 1} also exhibit poor orientation indexing. However, the indexing improves for Kikuchi patterns captured at 1 ms and 0.5 ms exposure time using our trained conditional generative diffusion model. See Figure \ref{fig:DDPM_restored IPF} and Figure \ref{fig:DDPM_restored_0.5ms 1}. Interestingly, the restored maps recover the grain structures and local orientations, closely resembling the ground truth.

\subsection{Restored patterns indexing metrics}
\begin{figure}[H]
    \centering
    \begin{subfigure}{0.45\linewidth} 
        \centering
        \includegraphics[width=\linewidth]{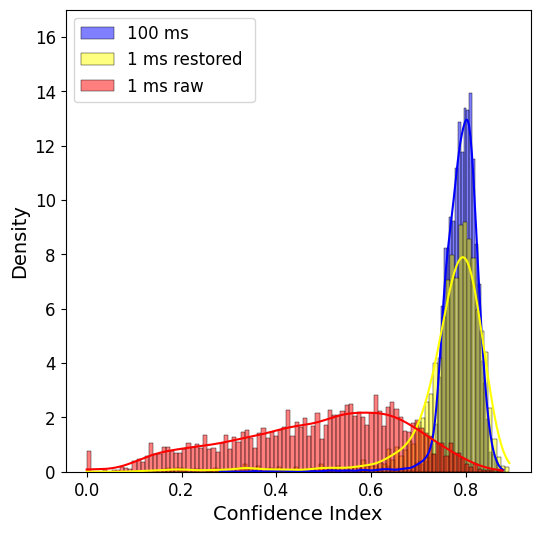}
        \caption{}
        \label{fig:cm_1ms}
    \end{subfigure}
    \hfill
    \begin{subfigure}{0.45\linewidth}
        \centering
        \includegraphics[width=\linewidth]{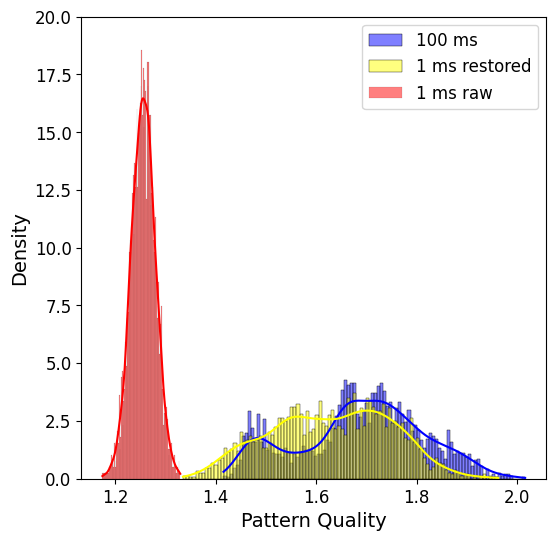}
        \caption{}
        \label{fig:pq_1ms}
    \end{subfigure}
    
    \caption{Comparative distribution plot of (a) CI of Hough indexing and (b) PQ metrics for all the indexed patterns, for comparing 100 ms patterns and restored 1 ms patterns, against raw 1 ms patterns.}
    \label{fig:IPF_comparison}
\end{figure}

\begin{figure}[H]
    \centering
    \begin{subfigure}{0.45\linewidth} 
        \centering
        \includegraphics[width=\linewidth]{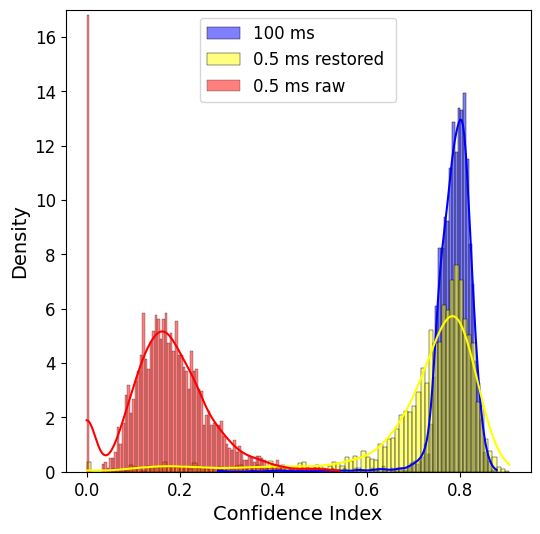}
        \caption{}
        \label{fig:cm_point5ms}
    \end{subfigure}
    \hfill
    \begin{subfigure}{0.45\linewidth}
        \centering
        \includegraphics[width=\linewidth]{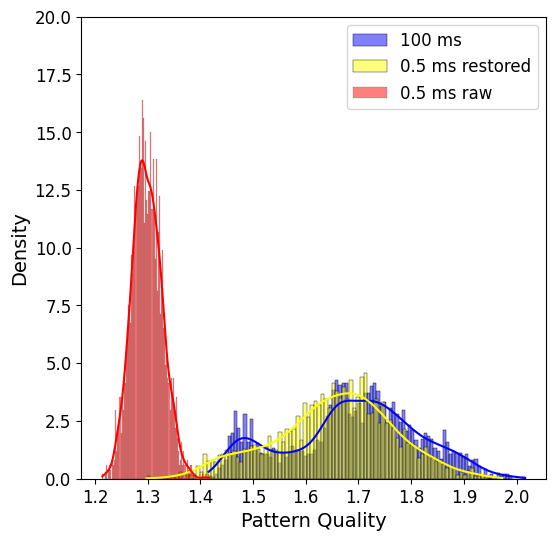}
        \caption{}
        \label{fig:pq_point5ms}
    \end{subfigure}
    
    \caption{Comparative distribution plot of (a) CI of Hough indexing and (b) PQ metrics for all the patterns in the scan map, for comparing 100 ms patterns and restored 0.5 ms patterns, against raw 0.5 ms patterns.}
    \label{fig:IPF_comparison}
\end{figure}

\begin{figure}[H]
    \centering
    
    \begin{subfigure}{1\linewidth}
        \centering
        \includegraphics[width=1\linewidth]{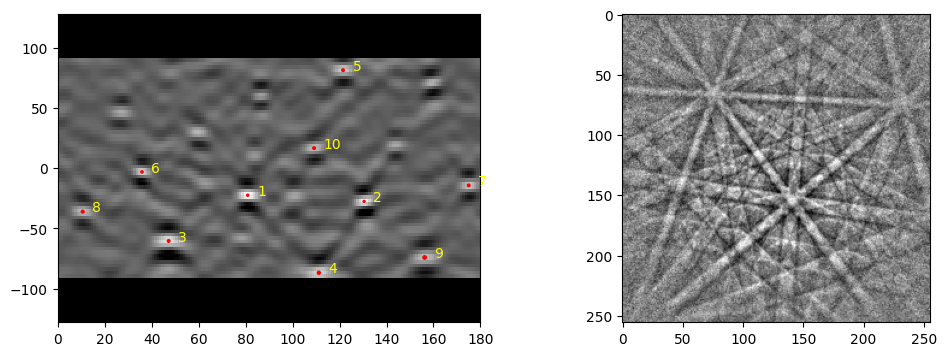}
        \caption{}
        \label{}
    \end{subfigure}
    \hfill
    \begin{subfigure}{1\linewidth}
        \centering
        \includegraphics[width=1\linewidth]{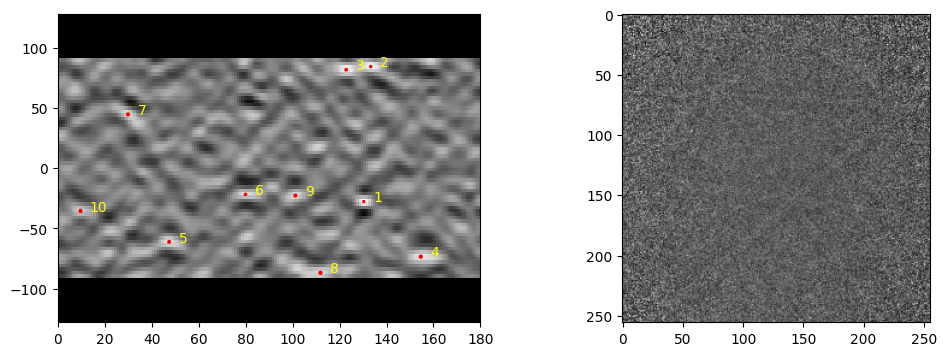}
        \caption{}
        \label{}
    \end{subfigure}
    \hfill
    \begin{subfigure}{1\linewidth} 
        \includegraphics[width=1\linewidth]{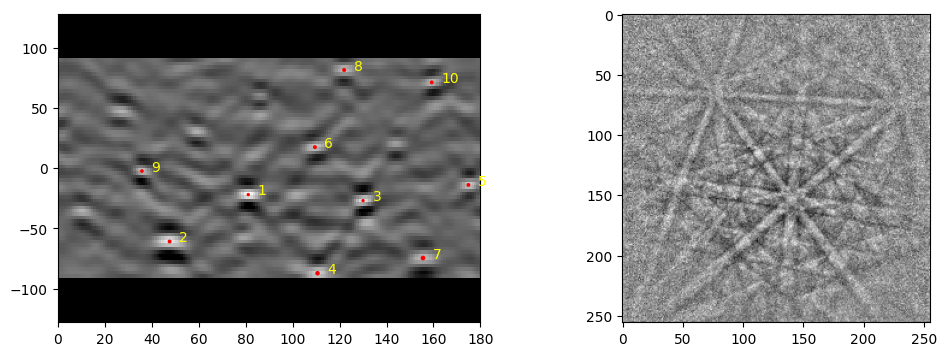}
        \caption{}
        \label{}
    \end{subfigure}
    \caption{Points identified in Hough space for (a) 100 ms patterns for benchmarking, (b) 1 ms raw experimental pattern and (c) restored 1 ms patterns: processed 1 ms patterns helps in better identification of lines and poles than the raw 1 ms patterns for Hough-based orientation indexing.}
    \label{fig:Restored Hough space}
\end{figure}

A quantitative assessment of the enhanced EBSD patterns was conducted to evaluate the improvement in orientation indexing performance. Two key metrics were used for this analysis - the Confidence Index (CI) \cite{field1997recent} and \textit{Pattern Quality (PQ)}. The CI values for an EBSD pattern indexing quantifies the reliability of the orientation indexing results. Higher CI values indicate greater confidence in the indexed orientations. The PQ values measures the sharpness and contrast of Kikuchi bands in a pattern. Higher PQ indicates clearer band features and better pattern clarity, aiding in more accurate indexing.
The evaluation of these indexing quality metrics demonstrates a significant improvement compared to the raw EBSD patterns. As shown in Figure \ref{fig:cm_1ms}, the CIs for most of the 1 ms exposure time restored patterns range between 0.6 and 0.9, with a mean of 0.75. This marks a substantial enhancement over the raw 1 ms patterns, which exhibited a mean CI of only 0.43. The improved CI indicates higher pattern quality and increased reliability in the orientation indexing process, approaching the confidence level of the highest quality 100 ms patterns captured experimentally, which had a mean CI of 0.8. Additionally, the improvement in the distribution of PQ for restored 1 ms patterns can also be seen in the Figure \ref{fig:pq_1ms}, which indicates improved sharpness in Kikuchi bands and makes them easily identifiable in the Hough space. Similarly, after processing 0.5 ms patterns the average CI of orientation indexing is 0.64, improved from the CI score of 0.2 for raw 0.5 ms patterns, as shown in Figure \ref{fig:cm_point5ms} and the corresponding improvement in the distribution of PQ is shown in the Figure \ref{fig:pq_point5ms}. The points identified in the Hough space for restored 1 ms patterns are more distinct and well defined than in the previous case of raw 1 ms patterns, as demonstrated by Figure \ref{fig:Restored Hough space}, which also shows the Hough space indexing for 100 ms Kikuchi patterns. 

\subsubsection{Predictions by diffusion model trained using low-resolution images of Kikuchi patterns}

\begin{figure}[H]
    \centering
    \includegraphics[width=0.4\linewidth]{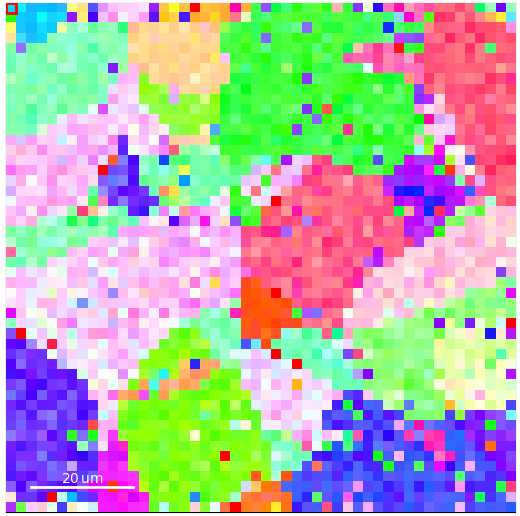}
    \caption{Orientation map for low-resolution (128 $\times$ 128), restored pattern from 1 ms exposure time. Suboptimal results after Hough based orientation indexing, due to poor resolution of training patterns.}
    \label{fig:low-res IPF}
\end{figure}

\begin{figure}[H]
    \centering
    \begin{subfigure}{0.45\linewidth} 
        \centering
        \includegraphics[width=\linewidth]{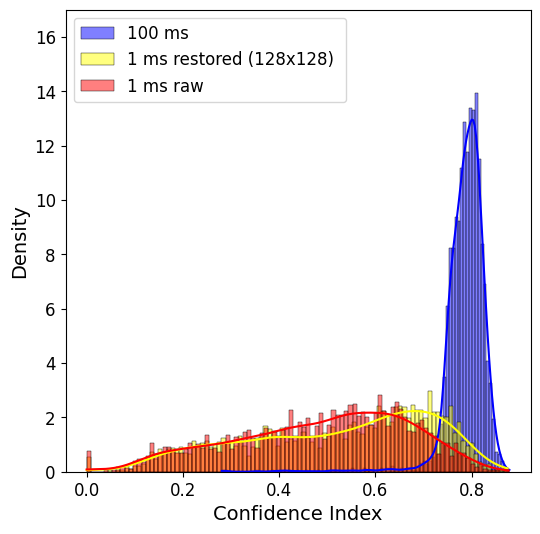}
        \caption{}
        \label{fig:cm_low_res}
    \end{subfigure}
    \hfill
    \begin{subfigure}{0.45\linewidth}
        \centering
        \includegraphics[width=\linewidth]{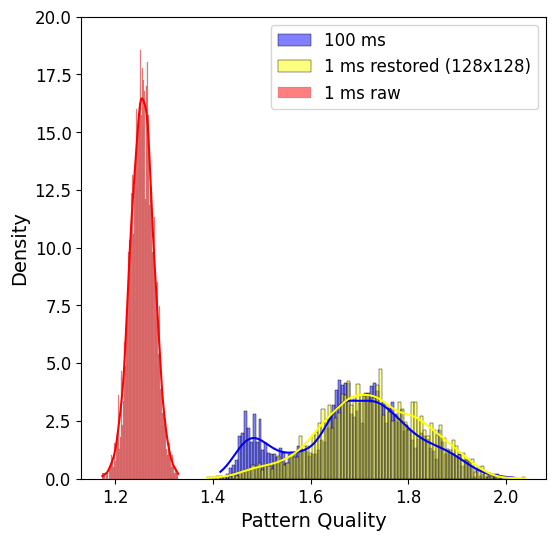}
        \caption{}
        \label{fig:pq_low_res}
    \end{subfigure}
    
    \caption{Comparative distribution plot of (a) CI of Hough indexing and (b) PQ metrics for restored 1 ms patterns trained at lower resolution (128 $\times$ 128), against raw 1 ms patterns. CI and PQ for 100 ms patterns is also shown for benchmarking. Pattern quality scores have been improved as compared to the scores for raw 1 ms patterns. However, the CI distribution shows that most of the indexed points lay on the lower confidence region, as such the restoration results are not very reliable.}
    \label{fig:cm_pq_low_res}
\end{figure}

\begin{figure}[H]
    \centering
    \includegraphics[width=1\linewidth]{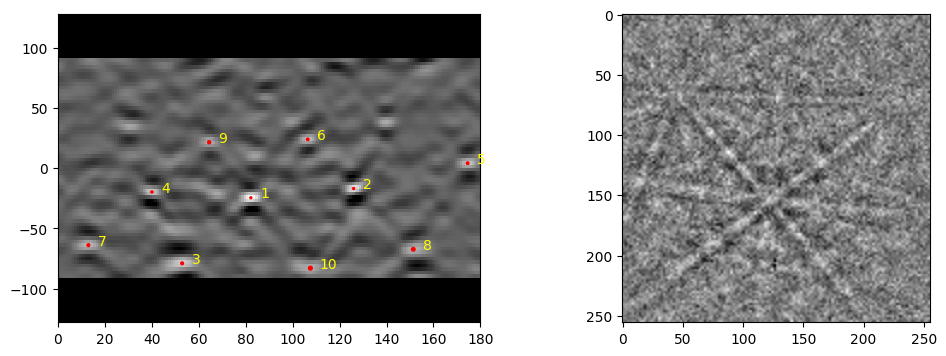}
    \caption{Points identified in Hough space for 1 ms restored pattern trained on low-resolution (128 $\times$ 128). }
    \label{fig:low-resHough}
\end{figure}

\noindent \noindent In addition to training the model on original-resolution patterns (i.e, 256 $\times$ 256), we also trained it on low-resolution patterns (128 $\times$ 128) to evaluate whether a reduced-order model could improve indexing accuracy after restoration. Using lower resolution patterns allows the diffusion model to operate more efficiently, requiring fewer parameters and enabling faster processing. However, our experiments revealed that while the restored patterns appear visually improved, they exhibit frequent distortions, artifacts such as curved Kikuchi bands. These distortions persist even when training the model for more epochs. Also doing so risks overfitting and potentially degrading the results further. Due to this issue, as evident in the Figures \ref{fig:low-res IPF} and \ref{fig:cm_pq_low_res}, the method struggles to accurately index them. Figure \ref{fig:low-resHough} shows an example Hough indexing for this restoration.

\section{Optimization of diffusion models for accelerated training and sampling}
\subsection{Implicit sampling in diffusion models}

\noindent While the DDPM-inspired model yielded enhanced EBSD patterns with high indexing accuracy, its training and inference processes were computationally expensive and time-consuming. Recent advancements in diffusion-based generative modeling have addressed this inefficiency by introducing techniques that accelerate both training and sampling, thereby making these models more feasible for high-throughput applications.

One such advancement is the implicit sampling algorithm, which provides a deterministic alternative to the stochastic sampling mechanism of DDPM. Traditional DDPM relies on a sequential noise removal process, where random noise is injected at each reverse diffusion step for higher stochasticity and diversity, resulting in slow and resource-heavy inference. In contrast, Denoising Diffusion Implicit Models (DDIM) \cite{song2020denoising} eliminate the need for stochastic sampling by deterministically defining the trajectory from the noise to the data, enabling faster generation while maintaining sample quality.

Let $\mathbf{x_T} \sim \mathcal{N}(0, I)$ denote a sample from the standard Gaussian prior, and let $\epsilon_\theta(\mathbf{x_t}, \mathbf{y}, t)$ be the model's prediction of noise at time step $t$. The standard DDPM reverse diffusion process is defined as a Markov chain \cite{ho2020denoising} as given by Eq. \ref{eq:ddpm}, DDIM replaces this stochastic transition with a deterministic update \cite{song2020denoising} as given by

\begin{equation}
\mathbf{x_{t-1}} = \sqrt{\bar{\alpha}_{t-1}} \left( \frac{\mathbf{x_t} - \sqrt{1 - \bar{\alpha}_t} \, \epsilon_\theta(\mathbf{x_t},\mathbf{y}, t)}{\sqrt{\bar{\alpha}_t}} \right) + \sqrt{1 - \bar{\alpha}_{t-1}} \cdot \epsilon_\theta(\mathbf{x_t}, \mathbf{y},t).
\label{eq:ddim}
\end{equation}

In this formulation, the first term represents the model’s estimate of the final clean image, derived from the current noisy input $\mathbf{x_t}$. The second term provides a direction of refinement, guiding the trajectory of the sample through time. This reformulation allows the model to traverse a deterministic path in data space, effectively reducing the number of required denoising steps. Since this reformulation is for the sampling process, the training objective of DDIM remains the same as that of DDPM.
In our experiments, we observed that the sampling process could be reduced from $T=500$ steps in DDPM to just $T=20$ steps using DDIM, with negligible degradation in the quality of the restored EBSD patterns. The number of steps were decided based on sampling time per pattern and the quality of pattern generation. This significant acceleration highlights the practical utility of implicit sampling for real-time or large-scale applications. 

\subsection{Latent space training and sampling in diffusion models}

While implicit sampling significantly accelerates the inference, it does not directly address the high computational cost associated with training diffusion models in high-dimensional pixel space (e.g., $256 \times 256 $ pixels). To overcome this, Latent Diffusion Models (LDMs) \cite{rombach2022high} introduce a more scalable framework by performing the diffusion process in a compressed, lower-dimensional latent space rather than in pixel space.

\begin{figure}[H]
    \centering
    \includegraphics[width=0.85\linewidth]{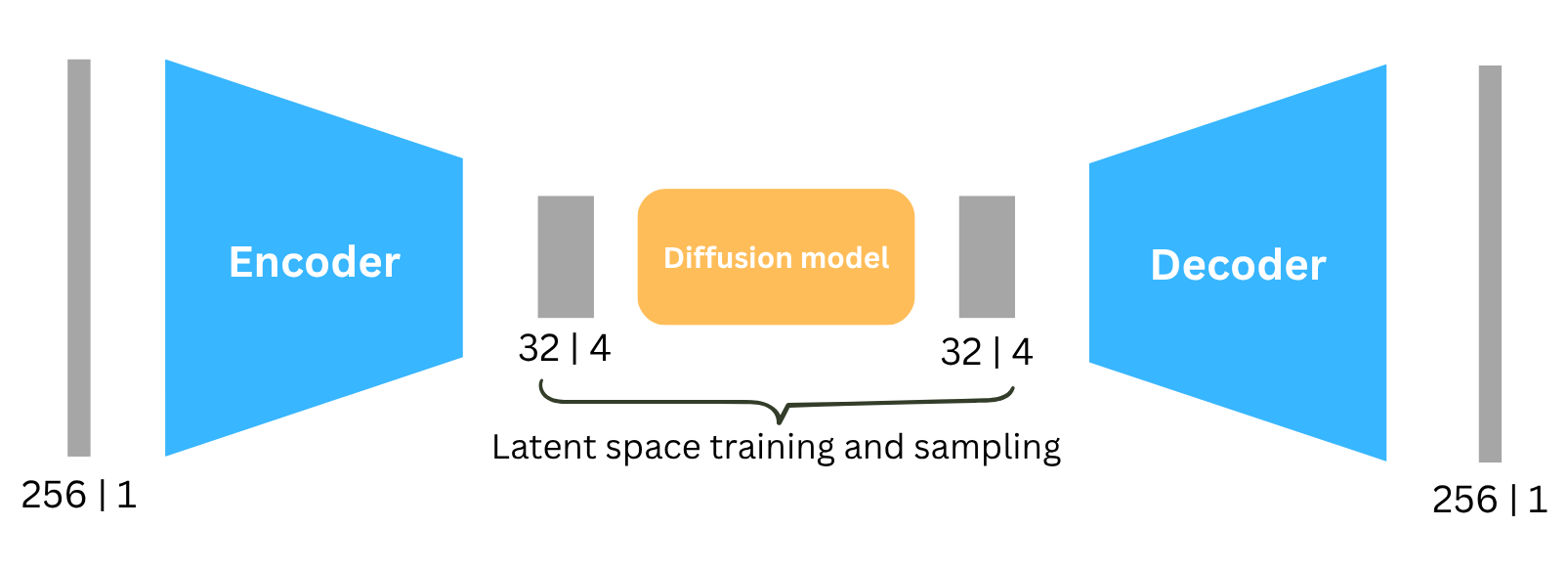}
    \caption{Latent space diffusion modelling framework.}
    \label{fig:LDM_model}
\end{figure}

Inspired by that approach we leveraged the stable diffusion's pretrained autoencoder \cite{rombach2022high} to encode high-resolution EBSD patterns into a compact and semantically rich latent representation (e.g., $32 \times 32 \times 4 $ pixels). Diffusion model is then applied within this latent space, substantially reducing both training and sampling costs. After the denoising process is complete, the latent representation is decoded back into the image space to recover the enhanced pattern.

\begin{equation}
\mathbf{z} = \mathcal{E}(\mathbf{x}), \quad \mathbf{x} \approx \mathcal{D}(\mathbf{z}),
\end{equation}

where $\mathcal{E}$ is the encoder and $\mathcal{D}$ is the decoder. This autoencoder has been trained on large variety of data and it ensures that the latent representation $\mathbf{z}$ captures essential semantic and structural features of the image $\mathbf{x}$. The overview of LDM framework is shown in Figure \ref{fig:LDM_model}.

The LDM framework with the implicit sampling algorithm thus provides a twofold advantage: (1) reduced memory and compute requirements due to the lower dimensionality of the latent space, and (2) faster convergence and inference due to reduced sampling steps and low memory requirements, all while preserving the fidelity of the restored EBSD patterns. 

We scaled down our original DDPM inspired architecture to handle the compact semantically rich latent space encoding of training patterns having the shape of $32 \times 32 \times 4 $. Moreover we removed a down sampling layer and a self attention layer to make the U-Net architecture lighter. This modified Latent Diffusion Model (mLDM) architecture follows the similar training process with the hyperparameters, and the total number of trainable parameters of the model are 5,617,636 which is less than the total of 5,881,569 parameters used earlier. In our implementation, we observed a significant reduction in training time and GPU memory usage, making latent space diffusion model with implicit sampling a practical solution for high-throughput EBSD pattern enhancement.

\subsection{Results and analysis}
\subsubsection{Restored pattern quality and indexing accuracy}
\noindent Once the model is trained in the latent space, the diffusion process is executed entirely within this lower-dimensional representation. During sampling, the raw noisy EBSD pattern is first encoded into the latent space using the pre-trained encoder and this encoding is used as the condition which is concatenated with the random noise to initiate the generation process. The denoising steps are guided by this condition, and the diffusion model iteratively refines the latent representation toward a high-quality reconstruction.

\begin{figure}[H]
    \centering
    \begin{subfigure}{0.18\linewidth}
        \centering
        \includegraphics[width=\linewidth]{image_3.png}
    \end{subfigure}
    \begin{subfigure}{0.18\linewidth}
        \centering
        \includegraphics[width=\linewidth]{1ms_image_3.png}
    \end{subfigure}
    \begin{subfigure}{0.18\linewidth}
        \centering
        \includegraphics[width=\linewidth]{point5ms_image_3.png}
    \end{subfigure}
    \begin{subfigure}{0.18\linewidth}
        \centering
        \includegraphics[width=\linewidth]{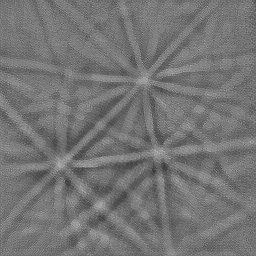}
    \end{subfigure}
        \begin{subfigure}{0.18\linewidth}
        \centering
        \includegraphics[width=\linewidth]{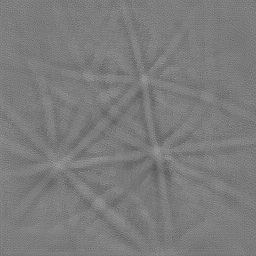}
    \end{subfigure}

    \begin{subfigure}{0.18\linewidth}
        \centering
        \includegraphics[width=\linewidth]{image_23.png}
    \end{subfigure}
    \begin{subfigure}{0.18\linewidth}
        \centering
        \includegraphics[width=\linewidth]{1ms_image_23.png}
    \end{subfigure}
    \begin{subfigure}{0.18\linewidth}
        \centering
        \includegraphics[width=\linewidth]{point5ms_image_23.png}
    \end{subfigure}
    \begin{subfigure}{0.18\linewidth}
        \centering
        \includegraphics[width=\linewidth]{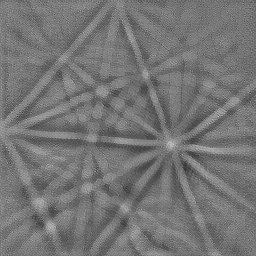}
    \end{subfigure}
        \begin{subfigure}{0.18\linewidth}
        \centering
        \includegraphics[width=\linewidth]{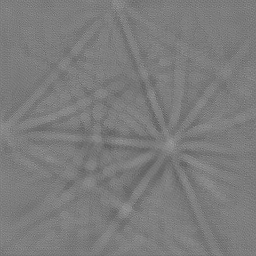}
    \end{subfigure}

    \begin{subfigure}{0.18\linewidth}
        \centering
        \includegraphics[width=\linewidth]{image_30.png}
        \caption{}
    \end{subfigure}
    \begin{subfigure}{0.18\linewidth}
        \centering  
        \includegraphics[width=\linewidth]{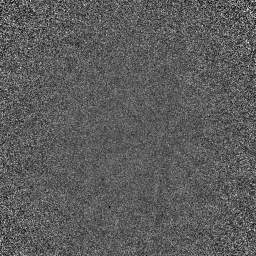}
        \caption{}
    \end{subfigure}
    \begin{subfigure}{0.18\linewidth}
        \centering
        \includegraphics[width=\linewidth]{point5ms_image_760.png}
        \caption{}
    \end{subfigure}
    \begin{subfigure}{0.18\linewidth}
        \centering
        \includegraphics[width=\linewidth]{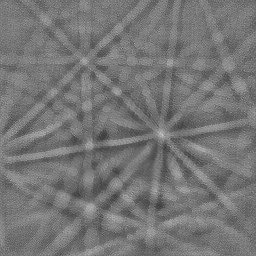}
        \caption{}
    \end{subfigure}
    \begin{subfigure}{0.18\linewidth}
        \centering
        \includegraphics[width=\linewidth]{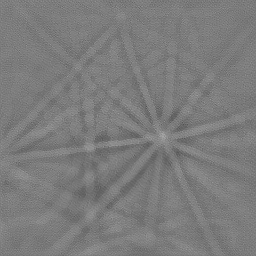}
        \caption{}
    \end{subfigure}
    \caption{Kikuchi patterns enhanced by the mLDM. (a) The Kikuchi patterns captured at 100 ms using the experiment (ground truth). (b)  Kikuchi patterns experimentally captured at 1 ms exposure time (or raw 1 ms patterns). (c) Kikuchi patterns experimentally captured at 0.5 ms exposure time (or raw 0.5 ms patterns). (d) 1 ms patterns restored by mLDM. (e) 0.5 ms patterns restored by mLDM.}
    
    \label{fig:diffusion_collage_LDM}
\end{figure}

\begin{figure}[H]
    \centering
    \begin{subfigure}{0.3\linewidth}
        \centering
        \includegraphics[width=\linewidth]{100ms_HI_orientation_map.png}
        \caption{}
        \label{fig:IPF_100ms 2}
    \end{subfigure}
    \hfill
    \begin{subfigure}{0.3\linewidth}
        \centering
        \includegraphics[width=\linewidth]{1ms_Patterns_HI_orientation_map.png}
        \caption{}
        \label{fig:IPF 1ms}
    \end{subfigure}
    \hfill
    \begin{subfigure}{0.3\linewidth}
        \centering
        \includegraphics[width=\linewidth]{0.5ms_Patterns_HI.png}
        \caption{}
        \label{fig:raw_0.5ms 1}
    \end{subfigure}

    \vspace{1em} 

    \begin{subfigure}{0.3\linewidth}
        \centering
        \includegraphics[width=\linewidth]{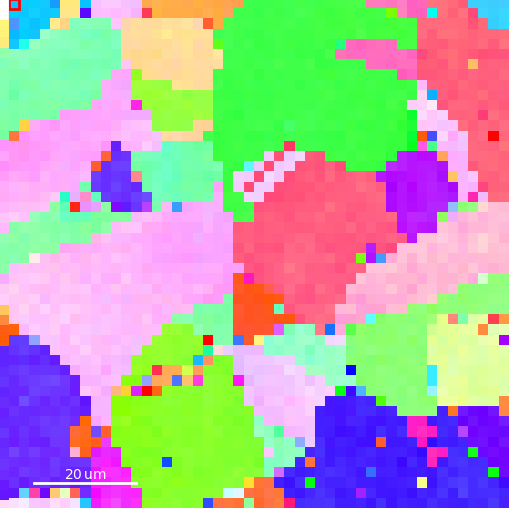}
        \caption{}
        \label{fig:restored IPF}
    \end{subfigure}
    \hspace{1em}
    \begin{subfigure}{0.3\linewidth}
        \centering
        \includegraphics[width=\linewidth]{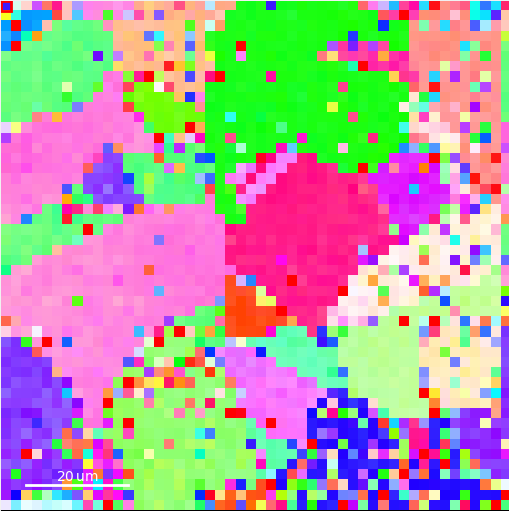}
        \caption{}
        \label{fig:restored_0.5ms 1}
    \end{subfigure}

    \caption{Comparison of orientation map for patterns captured at 100 ms, 1 ms and 0.5 ms exposure times with the orientation map of restored patterns by mLDM. Orientation map obtained by Hough indexing of experimentally captured pattern at (a) 100 ms exposure time, (b) 1 ms exposure time, (c) 0.5 ms exposure time. Orientation map based on restored Kikuchi patterns for (d) 1 ms exposure time (e) 0.5 ms exposure time. Significant improvement in the orientation indexing of restored Kikuchi patterns as compared to the orientation map of raw 1 ms exposure time patterns after passing them for processing through our trained mLDM. However, the restoration of 0.5 ms patterns through the mLDM faced some limitations because of the information loss during the latent space encoding process.}
    \label{fig:IPF_comparison1}
    
\end{figure}

After completion of the diffusion process, the final denoised latent vector is passed to the decoder to obtain the restored EBSD pattern in the original image space with dimensions $256 \times 256$. A selection of restored patterns generated by the mLDM is presented in Figure~\ref{fig:diffusion_collage_LDM}, demonstrating the model's ability to recover fine details of Kikuchi patterns from severely degraded input and enhancing the final orientation map upon performing orientation indexing on processed patterns, see Figure \ref{fig:IPF_comparison1}.

\subsection{Quantitative evaluation of orientation indexing enhancement}

\begin{figure}[H]
    \centering
    \begin{subfigure}{0.45\linewidth} 
        \centering
        \includegraphics[width=\linewidth]{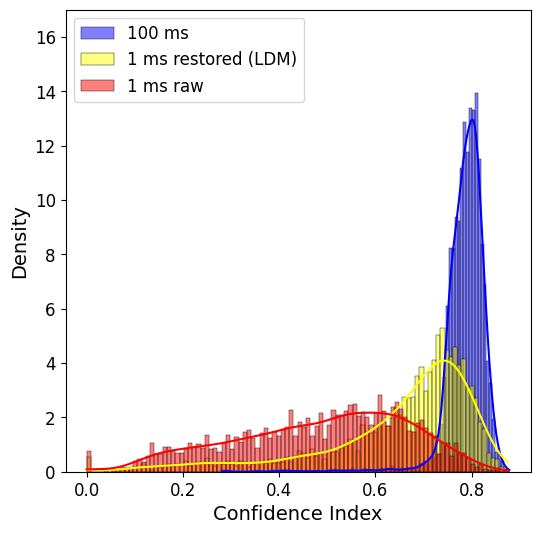}
        \caption{}
        \label{fig:LDM_CM}
    \end{subfigure}
    \hfill
    \begin{subfigure}{0.45\linewidth}
        \centering
        \includegraphics[width=\linewidth]{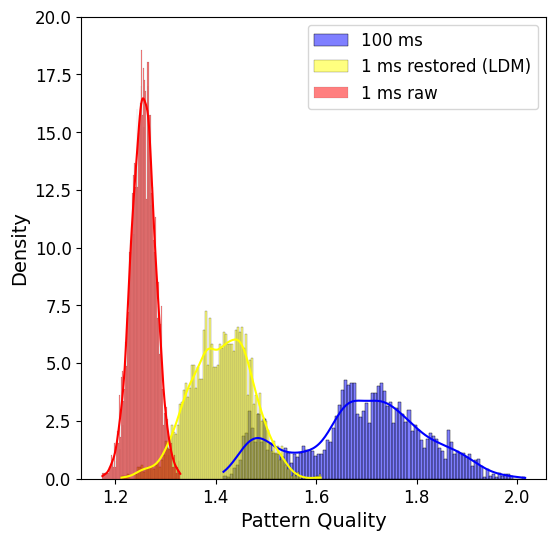}
        \caption{}
        \label{fig:LDM_PQ}
    \end{subfigure}
    
    \caption{Comparative distribution plot of (a) CI of Hough indexing and (b) PQ metrics for all the patterns in the scan map, for comparing 100 ms patterns and restored 1 ms patterns by mLDM, against raw 1 ms patterns.} 
    \label{fig:LDM_CM_and_PQ}
\end{figure}

\noindent A quantitative assessment of the enhanced EBSD patterns was conducted to evaluate the improvement in orientation indexing performance by analyzing CI and PQ. The distributions of these metrics for all indexed patterns are shown in Figures~\ref{fig:LDM_CM} and~\ref{fig:LDM_PQ}, respectively.

\noindent The evaluation results reveal a significant enhancement in indexing quality compared to the original low-exposure patterns. As illustrated by the distribution of CI in Figure~\ref{fig:LDM_CM}, the overall CI improves, with the enhanced patterns achieving a mean CI of 0.65 as compared to mean CI of raw 1 ms pattern being 0.43.  However, this score is suboptimal as compared to the score of 0.75 which was obtained by working directly on the pixel space of dimension 256 $\times$ 256. The reason for this is the loss occurring in the encoding of noisy input conditional patterns, which is confirmed by the fact that the orientation indexing performed by decoding these encoded conditional 1 ms patterns provides a lower average CI of 0.22. Therefore, the improved score of 0.65 is a substantial improvement even after the information loss in the latent space.

Similarly, the pattern quality metric shows a marked improvement, as depicted in Figure~\ref{fig:LDM_PQ}. The distribution shift toward higher PQ values indicates enhanced sharpness and clarity of Kikuchi bands, which facilitates more reliable detection and indexing in the Hough transform space. 

\subsection{Epoch-wise convergence and pattern quality improvement}
\begin{figure}[H]
    \centering
    
    \begin{subfigure}[b]{0.49\textwidth}
        \centering
        \includegraphics[trim={0 0 0 2cm},clip,width=\linewidth]{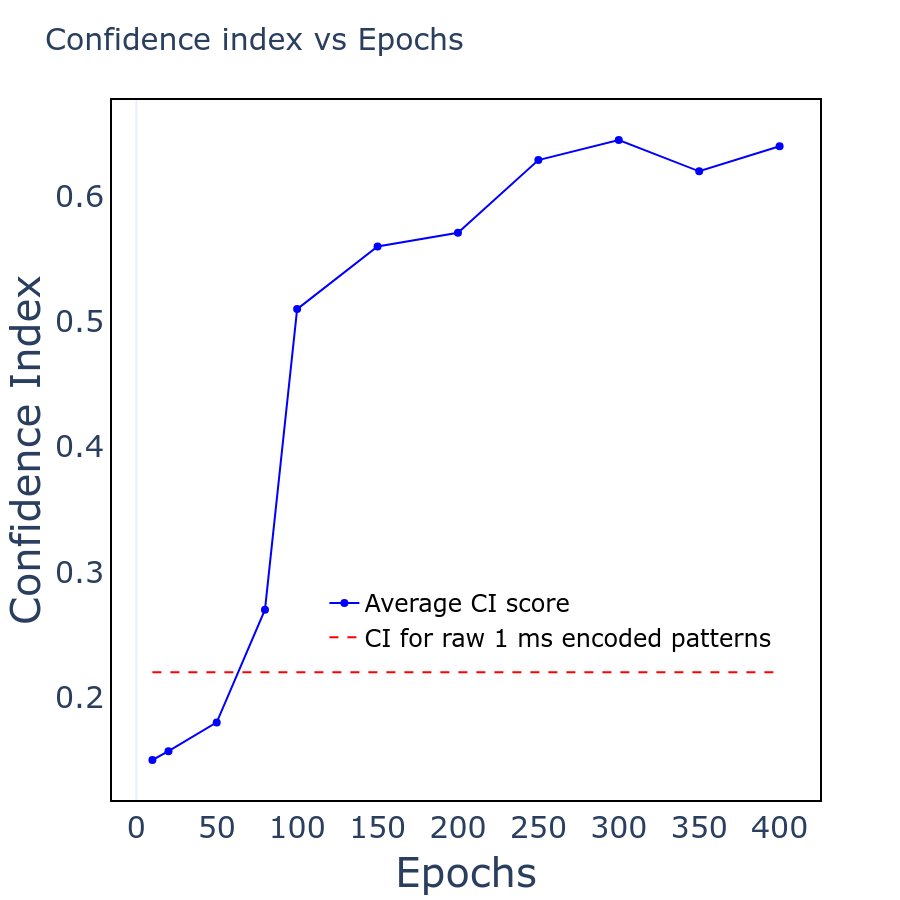}
        \caption{}
        \label{fig:raw_0.5ms}
    \end{subfigure}
    \hfill
    \begin{subfigure}[b]{0.49\textwidth}
        \centering
        \includegraphics[trim={0 0 0 2cm},clip,width=\linewidth]{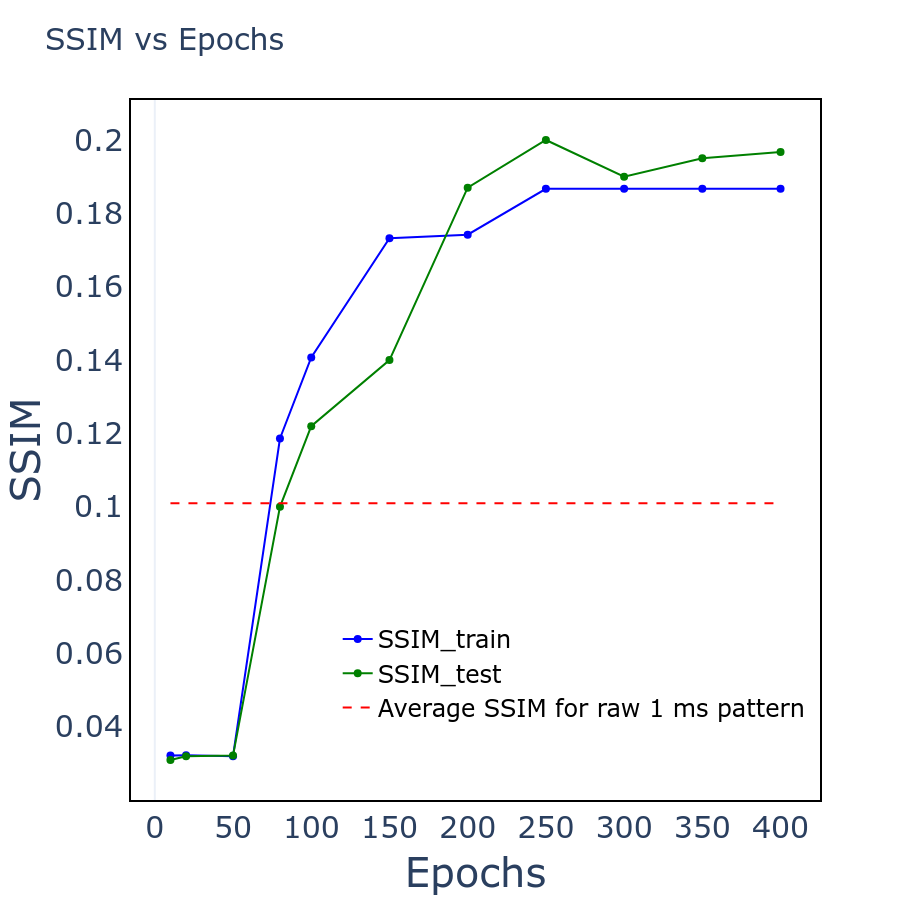}
        \caption{}
        \label{fig:restored_0.5ms}
    \end{subfigure}
    \begin{subfigure}[b]{0.49\textwidth}
        \centering
        \includegraphics[trim={0 0 0 2cm},clip,width=\linewidth]{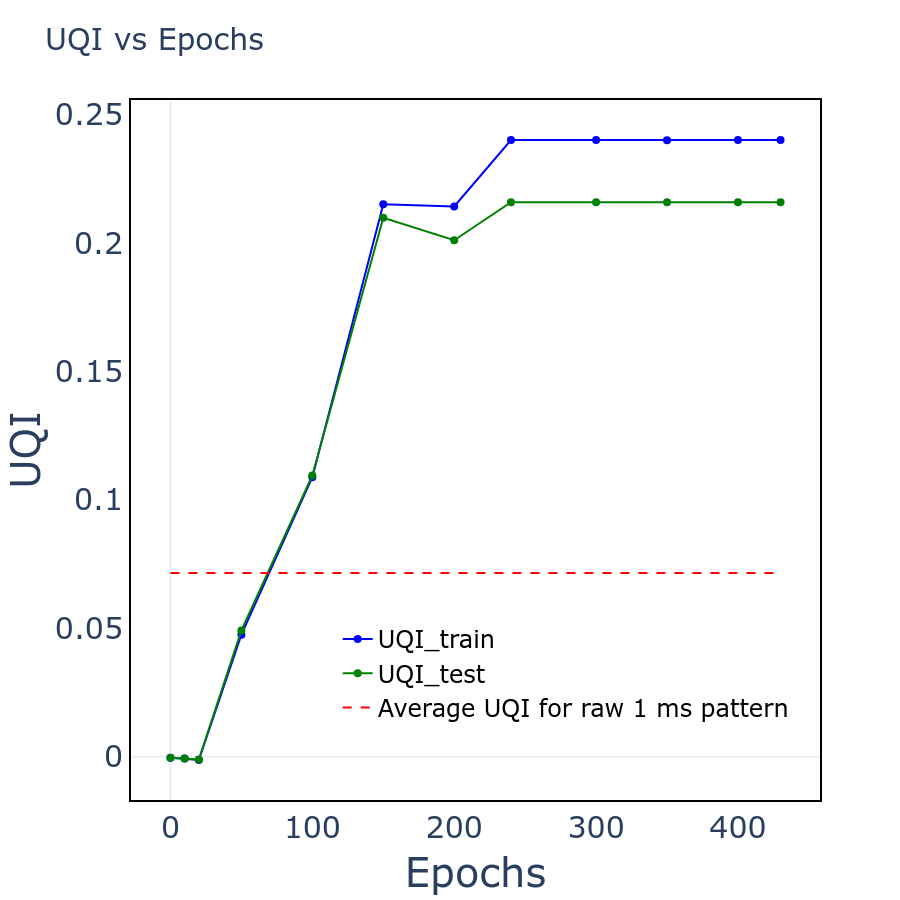}
        \caption{}
        \label{fig:raw_0.5ms}
    \end{subfigure}
    \hfill
    \begin{subfigure}[b]{0.49\textwidth}
        \centering
        \includegraphics[trim={0 0 0 2cm},clip,width=\linewidth]{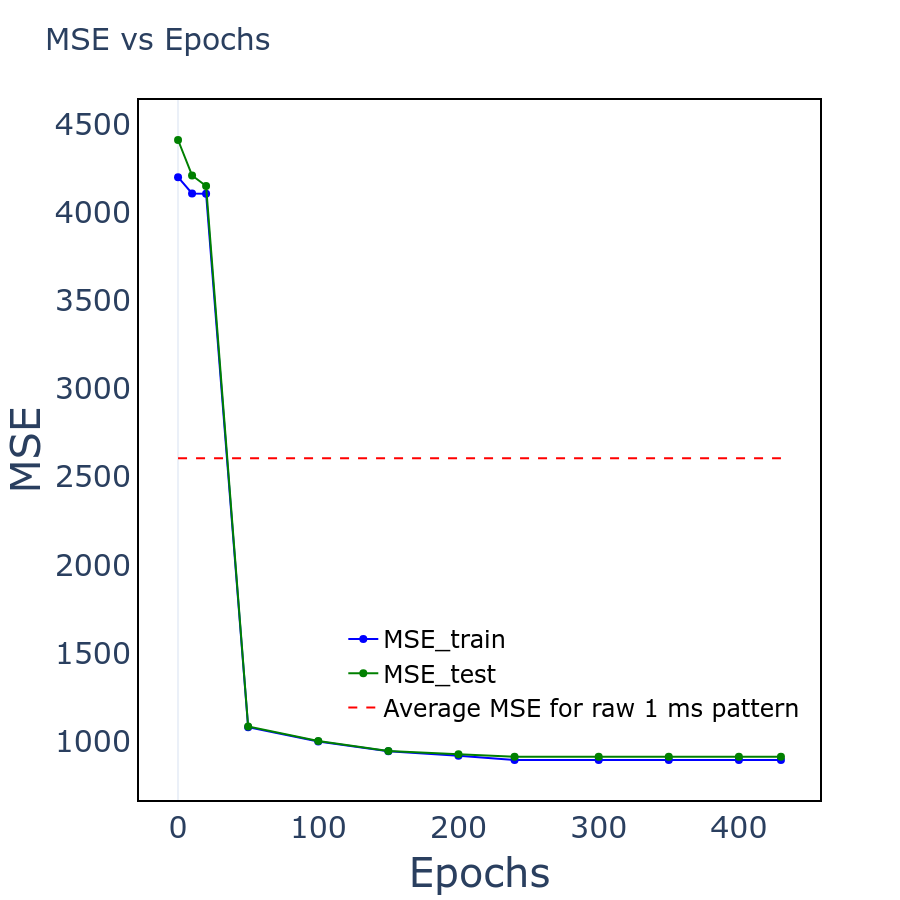}
        \caption{}
        \label{fig:restored_0.5ms}
    \end{subfigure}

  \caption{Evaluation of CI, SSIM, UQI and MSE, with the number of training epochs of the mLDM model. (a) Tradeoff of improvement in average CI of orientation indexing vs the number of epochs of training the mLDM, where the dashed red line shows the baseline CI (0.43) for the encoded raw 1 ms patterns after decoding. But while encoding, some information gets lost, so for raw encoded patterns after decoding the CI lowers down to 0.22. (b) Improvement in average SSIM between ground truth and restored patterns vs the number of epochs during training, while the dashed red line indicates baseline SSIM score between raw 1 ms patterns and ground truth i.e. 100 ms patterns. (c) Improvement in average UQI between the ground truth and restored patterns vs the number of epochs during training, the dashed red line indicates the baseline average UQI score. (d) Reduction in average MSE between the ground truth and restored patterns vs the number of epochs during training, the dashed red line indicates baseline average MSE between raw 1 ms patterns and ground truth.}
    \label{fig:LDM_metrices}
\end{figure}

\begin{figure}[H]
    \centering
    \includegraphics[width=1\linewidth]{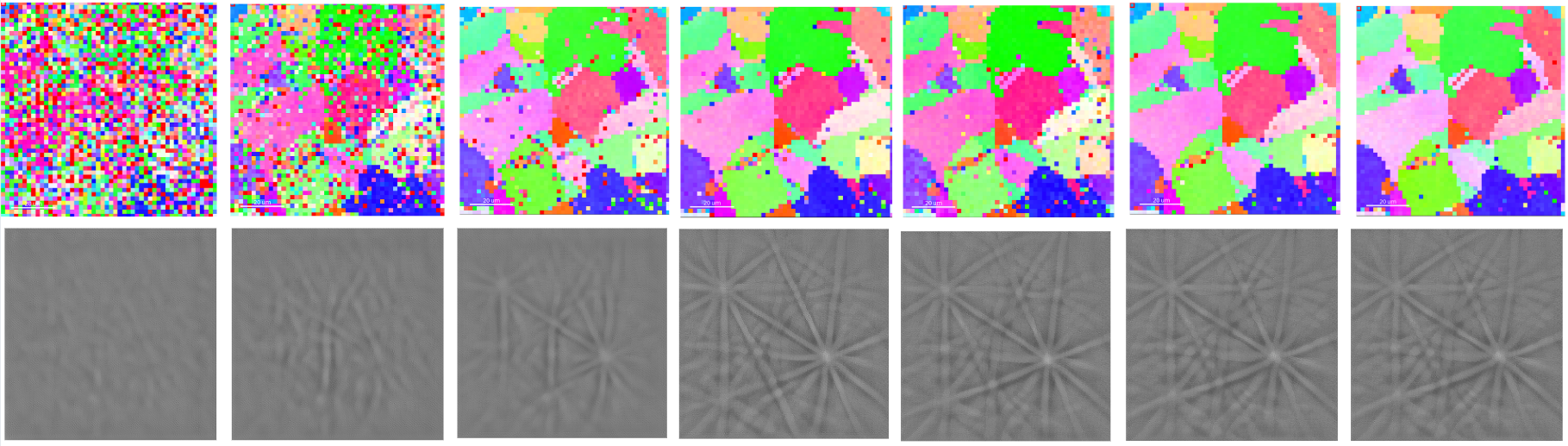}
    \caption{Improvement in orientation indexing with the number of epochs of training, demonstrated by the orientation map of restored patterns obtained at different epochs, i.e, 50, 100, 150, 200, 250, 300 and 350. A sample of a restored Kikuchi pattern corresponding to each epoch is shown below the respective orientation map.}
    \label{fig:epoch_wise_quality}
\end{figure}

\noindent Convergence analysis is important to assess the model's learning progress and evaluate the training stability. A detailed convergence analysis of the mLDM was performed using the following quantitative metrics: CI, Unique Quality Index (UQI), Structural Similarity Index Measure (SSIM), and Mean Squared Error (MSE). UQI considers correlation, luminance, and contrast for the comparison between the images. SSIM is an improvement over UQI which also considers perceptual similarity for comparison. MSE computes the average of the squared differences between corresponding pixel values of two images. Our model was trained for a total of 400 epochs. As shown in Figure~\ref{fig:LDM_metrices}, a consistent improvement across all four metrics is observed, with convergence in pattern quality and indexing performance occurring around epochs 250 to 300. To further illustrate this progression, Figure~\ref{fig:epoch_wise_quality} presents the epoch-wise visual enhancement of a representative EBSD pattern. Alongside, the corresponding orientation maps obtained from indexing the restored patterns at various training stages demonstrate the gradual improvement in orientation indexing accuracy. 

\subsection{Speed-up calculations}
\noindent To evaluate the computational efficiency of the mLDM framework over the baseline implementation inspired by conditional DDPM, we conducted a systematic study encompassing both the training and inference phases. The key metrics considered include epoch-wise training time, sampling time per pattern, batch-wise memory utilization, and total time required to enhance an entire EBSD scan map.

 \subsubsection{Training acceleration}
\noindent With the introduction of the mLDM, the training time per epoch was significantly reduced. In the original conditional DDPM setup, training took approximately 150 seconds per epoch with maximum GPU utilization i.e.\textasciitilde 75 GB usage of the total available memory of 80GB on the single A100 GPU that was used. In contrast, the mLDM, operating in the compressed latent space (i.e, 34 $\times$ 34 $\times$ 4) with a batch size of 64, reduced the epoch time to 1.5-2 seconds. Although convergence now typically requires 280–300 epochs compared to 150 epochs in the pixel-space (256 $\times$ 256) training of the DDPM. The drastic reduction in per epoch training time results in an \textasciitilde 45 times speedup in total training time till convergence. This acceleration is attributed to both the reduced dimensionality of the latent representation and the efficient optimization dynamics in the latent space.

\subsubsection{Sampling acceleration}
\begin{figure}[H]
    \centering
    \includegraphics[trim={0 0 0 2cm},clip,width=0.5\linewidth]{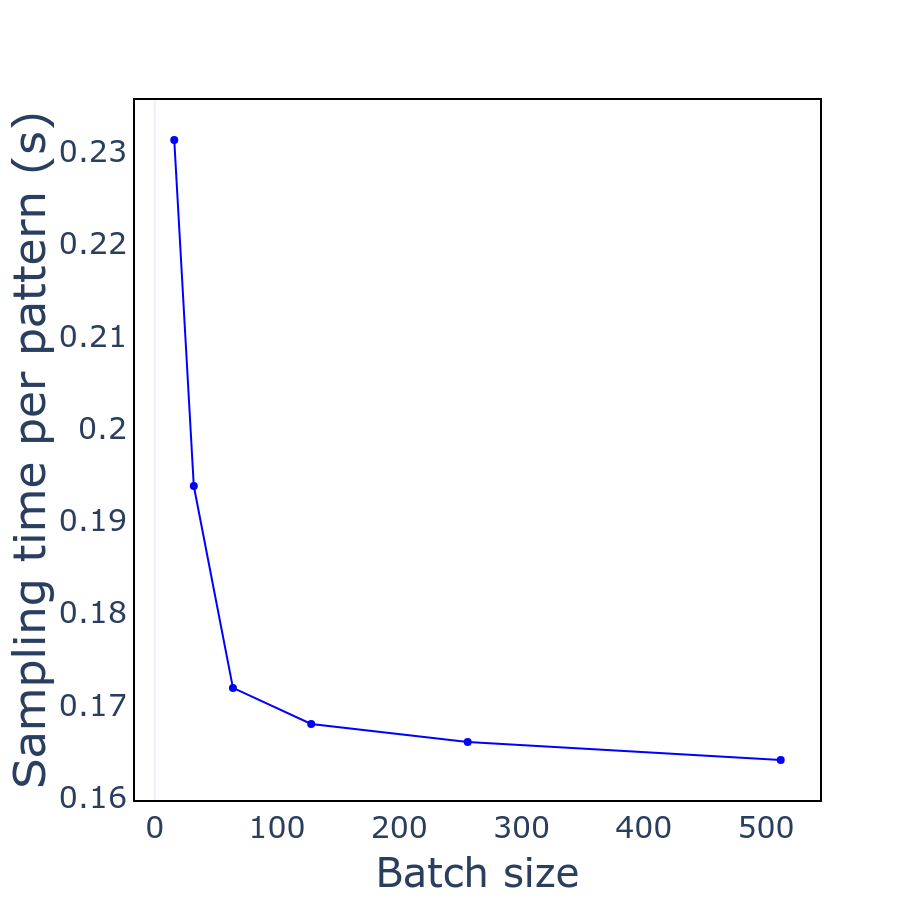}
    \caption{Sampling time per pattern tradeoff with the batch size used for sampling via the mLDM, indicating the saturation in speedup with the larger batch sizes.}
    \label{fig:time_vs_batch}
\end{figure}

\noindent Sampling time was evaluated by measuring the time taken to restore a single low-quality input Kikuchi pattern. In the baseline DDPM model, generating a high-quality pattern took 14-15 seconds per pattern. In contrast, the mLDM achieved a sampling time of \textasciitilde  0.165 seconds per pattern with an optimum batch size of 256, representing an \textasciitilde	 90$\times$ improvement in total sampling speed and \textasciitilde 275$\times$ faster diffusion sampling (excluding additional decoding time), which highlights the substantial computational benefit of operating in the latent space. Furthermore, scalability with the other batch sizes was also assessed. Figure \ref{fig:time_vs_batch} indicates saturation of sampling time per pattern as a function of batch size. The sampling time varied from 42s to 43s for various batches of size 256. Of this total sampling time, 14 seconds were spend on diffusion sampling and the remainder 29s for decoding via Stable Diffusion's pretrained decoder.

This analysis demonstrates that diffusion sampling in the mLDM framework is highly parallelizable and benefits substantially from increased batch sizes. To enhance the patterns of an entire EBSD scan map of 50$\times$50, the original DDPM setup required approximately 37,000 seconds. In contrast, with mLDM (batch size 256), the full scan can be sampled in \textasciitilde 400 seconds. 

\section{Evaluation metrics for various cases}

\noindent The evaluation of model performances trained for different cases are shown in Table \ref{tab:metrics}. Our model has shown improvement in the average SSIM, MSE, and CI scores for the restored patterns as compared to the raw patterns both at 1 ms and 0.5 ms exposure times. The average SSIM score between the raw patterns and 100 ms exposure time patterns have been found and compared with the average score between the restored patterns and 100 ms patterns. Also, the CI scores indicate a significant improvement in the quality of the orientation indexing for restored patterns through the diffusion models as shown in the Table \ref{tab:metrics}.

The results of our diffusion-based restoration frameworks underscore their potential to significantly improve the EBSD analysis. By effectively recovering weak bands and poles in low-quality Kikuchi patterns, the model enhances the reliability of traditional orientation indexing methods.


\begin{table}[H]
\centering
\caption{Evaluation of different metrics for various cases. All the results are corresponding to the patterns of the resolution of 256$\times$256, unless specified otherwise.}
\label{tab:metrics}
\resizebox{15cm}{!}{ 
\begin{tabular}{lccccc}
\hline
\textbf{Cases} & \textbf{CI} & \textbf{SSIM} & \textbf{MSE}  
\\ \hline
1 ms restored patterns (DDPM) & 0.75 & 0.2854 & 1189  \\ 
1 ms restored patterns (LDM)   & 0.65 & 0.2 & 913\\
1 ms raw encoded patterns & 0.22 & 0.101 & 2668  \\
1 ms raw patterns  & 0.43 & 0.113 & 2743 \\ 
0.5 ms restored patterns (DDPM)  & 0.64  & 0.2553 & 1277  \\ 
0.5 ms restored patterns (LDM)  & 0.3  & 0.1543 & 1033   \\
0.5 ms raw patterns  & 0.2 & 0.0835 & 4782    \\ 
1 ms low-res restored patterns (128 $\times$ 128)  & 0.51 & 0.1143 & 1761  \\ 
\hline
\end{tabular}
}

\end{table}


\section{Conclusion}
\noindent In this work, we developed conditional generative diffusion-based restoration models to enhance the quality of noisy Kikuchi patterns captured at short exposure times (0.5 ms and 1 ms). The proposed model is specifically designed to reconstruct and restore features lost due to reduced exposure, thereby generating high-quality Kikuchi patterns suitable for reliable post-processing in EBSD analyses. 
Our diffusion-based approach demonstrates strong restoration performance by successfully recovering many of the missing features in low-quality patterns. These features are often weak bands and poles in the Kikuchi pattern. These restored patterns can be effectively indexed using traditional Hough transform-based techniques, enabling accurate orientation determination with higher confidence indices.

Furthermore, to improve training efficiency and accelerate sampling, we adopted optimization strategies from the Denoising Diffusion Implicit Models (DDIM) and Latent Diffusion Models (LDM) frameworks. These frameworks accelerated both the training and sampling process significantly with low computational and memory requirements. However, they introduce some accuracy loss due to information degradation during latent space encoding by a universal pre-trained autoencoder. This issue can be mitigated by fine-tuning the autoencoder on EBSD data to better preserve relevant features in Kikuchi patterns.

Future work will focus on minimizing latent space information loss through domain-specific fine-tuning and further improving the model generalization by training on EBSD patterns captured under diverse experimental conditions, ensuring robust and reliable restoration performance across a wide range of orientations and experimental settings.

\section*{Acknowledgments}
\noindent ALR acknowledges the generous help from ThermoFisher Scientific Inc in providing the EBSD data of Ni. ALR also acknowledges the support from the IIT Bombay HPC Central Facility and the national mission GPU HPC facility Param Rudra.

\section*{Conflict of interest}

\noindent The authors have no relevant financial or non-financial interests to disclose.
	%

\bibliographystyle{elsarticle-num-names} 
\bibliography{references}

\begin{thebibliography}{41}
\expandafter\ifx\csname natexlab\endcsname\relax\def\natexlab#1{#1}\fi
\providecommand{\url}[1]{\texttt{#1}}
\providecommand{\href}[2]{#2}
\providecommand{\path}[1]{#1}
\providecommand{\DOIprefix}{doi:}
\providecommand{\ArXivprefix}{arXiv:}
\providecommand{\URLprefix}{URL: }
\providecommand{\Pubmedprefix}{pmid:}
\providecommand{\doi}[1]{\href{http://dx.doi.org/#1}{\path{#1}}}
\providecommand{\Pubmed}[1]{\href{pmid:#1}{\path{#1}}}
\providecommand{\bibinfo}[2]{#2}
\ifx\xfnm\relax \def\xfnm[#1]{\unskip,\space#1}\fi
\bibitem[{Schwartz et~al.(2009)Schwartz, Kumar, Adams, and
  Field}]{schwartz2009electron}
\bibinfo{author}{A.~J. Schwartz}, \bibinfo{author}{M.~Kumar},
  \bibinfo{author}{B.~L. Adams}, \bibinfo{author}{D.~P. Field},
  \bibinfo{title}{Electron backscatter diffraction in materials science},
  volume~\bibinfo{volume}{2}, \bibinfo{publisher}{Springer},
  \bibinfo{year}{2009}.
\bibitem[{Krishna et~al.(2023)Krishna, Madhavan, Pantawane, Banerjee, and
  Dahotre}]{krishna2023machine}
\bibinfo{author}{K.~M. Krishna}, \bibinfo{author}{R.~Madhavan},
  \bibinfo{author}{M.~V. Pantawane}, \bibinfo{author}{R.~Banerjee},
  \bibinfo{author}{N.~B. Dahotre},
\newblock \bibinfo{title}{Machine learning based de-noising of electron back
  scatter patterns of various crystallographic metallic materials fabricated
  using laser directed energy deposition},
\newblock \bibinfo{journal}{Ultramicroscopy} \bibinfo{volume}{247}
  (\bibinfo{year}{2023}) \bibinfo{pages}{113703}.
\bibitem[{Chen et~al.(2015)Chen, Park, Wei, Newstadt, Jackson, Simmons,
  De~Graef, and Hero}]{chen2015dictionary}
\bibinfo{author}{Y.~H. Chen}, \bibinfo{author}{S.~U. Park},
  \bibinfo{author}{D.~Wei}, \bibinfo{author}{G.~Newstadt},
  \bibinfo{author}{M.~A. Jackson}, \bibinfo{author}{J.~P. Simmons},
  \bibinfo{author}{M.~De~Graef}, \bibinfo{author}{A.~O. Hero},
\newblock \bibinfo{title}{A dictionary approach to electron backscatter
  diffraction indexing},
\newblock \bibinfo{journal}{Microscopy and Microanalysis} \bibinfo{volume}{21}
  (\bibinfo{year}{2015}) \bibinfo{pages}{739--752}.
\bibitem[{Lenthe et~al.(2019)Lenthe, Singh, and De~Graef}]{lenthe2019spherical}
\bibinfo{author}{W.~Lenthe}, \bibinfo{author}{S.~Singh},
  \bibinfo{author}{M.~De~Graef},
\newblock \bibinfo{title}{A spherical harmonic transform approach to the
  indexing of electron back-scattered diffraction patterns},
\newblock \bibinfo{journal}{Ultramicroscopy} \bibinfo{volume}{207}
  (\bibinfo{year}{2019}) \bibinfo{pages}{112841}.
\bibitem[{Ostormujof et~al.(2022)Ostormujof, Purohit, Breumier, Gey, Salib, and
  Germain}]{ostormujof2022deep}
\bibinfo{author}{T.~M. Ostormujof}, \bibinfo{author}{R.~P.~R. Purohit},
  \bibinfo{author}{S.~Breumier}, \bibinfo{author}{N.~Gey},
  \bibinfo{author}{M.~Salib}, \bibinfo{author}{L.~Germain},
\newblock \bibinfo{title}{Deep learning for automated phase segmentation in
  ebsd maps. a case study in dual phase steel microstructures},
\newblock \bibinfo{journal}{Materials Characterization} \bibinfo{volume}{184}
  (\bibinfo{year}{2022}) \bibinfo{pages}{111638}.
\bibitem[{Kaufmann et~al.(2021)Kaufmann, Lane, Liu, and
  Vecchio}]{kaufmann2021efficient}
\bibinfo{author}{K.~Kaufmann}, \bibinfo{author}{H.~Lane},
  \bibinfo{author}{X.~Liu}, \bibinfo{author}{K.~S. Vecchio},
\newblock \bibinfo{title}{Efficient few-shot machine learning for
  classification of ebsd patterns},
\newblock \bibinfo{journal}{Scientific reports} \bibinfo{volume}{11}
  (\bibinfo{year}{2021}) \bibinfo{pages}{8172}.
\bibitem[{Kaufmann et~al.(2020)Kaufmann, Zhu, Rosengarten, and
  Vecchio}]{kaufmann2020deep}
\bibinfo{author}{K.~Kaufmann}, \bibinfo{author}{C.~Zhu}, \bibinfo{author}{A.~S.
  Rosengarten}, \bibinfo{author}{K.~S. Vecchio},
\newblock \bibinfo{title}{Deep neural network enabled space group
  identification in ebsd},
\newblock \bibinfo{journal}{Microscopy and Microanalysis} \bibinfo{volume}{26}
  (\bibinfo{year}{2020}) \bibinfo{pages}{447--457}.
\bibitem[{Stoll and Benner(2021)}]{stoll2021machine}
\bibinfo{author}{A.~Stoll}, \bibinfo{author}{P.~Benner},
\newblock \bibinfo{title}{Machine learning for material characterization with
  an application for predicting mechanical properties},
\newblock \bibinfo{journal}{GAMM-Mitteilungen} \bibinfo{volume}{44}
  (\bibinfo{year}{2021}) \bibinfo{pages}{e202100003}.
\bibitem[{Choudhary et~al.(2022)Choudhary, DeCost, Chen, Jain, Tavazza, Cohn,
  Park, Choudhary, Agrawal, Billinge et~al.}]{choudhary2022recent}
\bibinfo{author}{K.~Choudhary}, \bibinfo{author}{B.~DeCost},
  \bibinfo{author}{C.~Chen}, \bibinfo{author}{A.~Jain},
  \bibinfo{author}{F.~Tavazza}, \bibinfo{author}{R.~Cohn},
  \bibinfo{author}{C.~W. Park}, \bibinfo{author}{A.~Choudhary},
  \bibinfo{author}{A.~Agrawal}, \bibinfo{author}{S.~J. Billinge}, et~al.,
\newblock \bibinfo{title}{Recent advances and applications of deep learning
  methods in materials science},
\newblock \bibinfo{journal}{npj Computational Materials} \bibinfo{volume}{8}
  (\bibinfo{year}{2022}) \bibinfo{pages}{59}.
\bibitem[{Andrews et~al.(2023)Andrews, Strantza, Calta, Matthews, and
  Taheri}]{andrews2023denoising}
\bibinfo{author}{C.~E. Andrews}, \bibinfo{author}{M.~Strantza},
  \bibinfo{author}{N.~P. Calta}, \bibinfo{author}{M.~J. Matthews},
  \bibinfo{author}{M.~L. Taheri},
\newblock \bibinfo{title}{A denoising autoencoder for improved kikuchi pattern
  quality and indexing in electron backscatter diffraction},
\newblock \bibinfo{journal}{Ultramicroscopy} \bibinfo{volume}{253}
  (\bibinfo{year}{2023}) \bibinfo{pages}{113810}.
\bibitem[{Shen et~al.(2019)Shen, Pokharel, Nizolek, Kumar, and
  Lookman}]{shen2019convolutional}
\bibinfo{author}{Y.-F. Shen}, \bibinfo{author}{R.~Pokharel},
  \bibinfo{author}{T.~J. Nizolek}, \bibinfo{author}{A.~Kumar},
  \bibinfo{author}{T.~Lookman},
\newblock \bibinfo{title}{Convolutional neural network-based method for
  real-time orientation indexing of measured electron backscatter diffraction
  patterns},
\newblock \bibinfo{journal}{Acta Materialia} \bibinfo{volume}{170}
  (\bibinfo{year}{2019}) \bibinfo{pages}{118--131}.
\bibitem[{Ding et~al.(2020)Ding, Pascal, and De~Graef}]{ding2020indexing}
\bibinfo{author}{Z.~Ding}, \bibinfo{author}{E.~Pascal},
  \bibinfo{author}{M.~De~Graef},
\newblock \bibinfo{title}{Indexing of electron back-scatter diffraction
  patterns using a convolutional neural network},
\newblock \bibinfo{journal}{Acta Materialia} \bibinfo{volume}{199}
  (\bibinfo{year}{2020}) \bibinfo{pages}{370--382}.
\bibitem[{Ding and Graef(2023)}]{Ding2023}
\bibinfo{author}{Z.~Ding}, \bibinfo{author}{M.~D. Graef},
\newblock \bibinfo{title}{Parametric simulation of electron backscatter
  diffraction patterns through generative models},
\newblock \bibinfo{journal}{npj Computational Materials} \bibinfo{volume}{9}
  (\bibinfo{year}{2023}) \bibinfo{pages}{199}. \URLprefix
  \url{https://doi.org/10.1038/s41524-023-01143-z}.
  \DOIprefix\doi{10.1038/s41524-023-01143-z}.
\bibitem[{Goodfellow et~al.(2014)Goodfellow, Pouget-Abadie, Mirza, Xu,
  Warde-Farley, Ozair, Courville, and Bengio}]{goodfellow2014generative}
\bibinfo{author}{I.~J. Goodfellow}, \bibinfo{author}{J.~Pouget-Abadie},
  \bibinfo{author}{M.~Mirza}, \bibinfo{author}{B.~Xu},
  \bibinfo{author}{D.~Warde-Farley}, \bibinfo{author}{S.~Ozair},
  \bibinfo{author}{A.~Courville}, \bibinfo{author}{Y.~Bengio},
\newblock \bibinfo{title}{Generative adversarial nets},
\newblock \bibinfo{journal}{Advances in neural information processing systems}
  \bibinfo{volume}{27} (\bibinfo{year}{2014}).
\bibitem[{Kingma et~al.(2013)Kingma, Welling et~al.}]{kingma2013auto}
\bibinfo{author}{D.~P. Kingma}, \bibinfo{author}{M.~Welling}, et~al.,
  \bibinfo{title}{Auto-encoding variational bayes}, \bibinfo{year}{2013}.
\bibitem[{Ho et~al.(2020)Ho, Jain, and Abbeel}]{ho2020denoising}
\bibinfo{author}{J.~Ho}, \bibinfo{author}{A.~Jain},
  \bibinfo{author}{P.~Abbeel},
\newblock \bibinfo{title}{Denoising diffusion probabilistic models},
\newblock \bibinfo{journal}{Advances in neural information processing systems}
  \bibinfo{volume}{33} (\bibinfo{year}{2020}) \bibinfo{pages}{6840--6851}.
\bibitem[{Wang et~al.(2023)Wang, Sun, Chehri, and Song}]{wang2023review}
\bibinfo{author}{X.~Wang}, \bibinfo{author}{L.~Sun},
  \bibinfo{author}{A.~Chehri}, \bibinfo{author}{Y.~Song},
\newblock \bibinfo{title}{A review of gan-based super-resolution reconstruction
  for optical remote sensing images},
\newblock \bibinfo{journal}{Remote Sensing} \bibinfo{volume}{15}
  (\bibinfo{year}{2023}) \bibinfo{pages}{5062}.
\bibitem[{Saharia et~al.(2022)Saharia, Ho, Chan, Salimans, Fleet, and
  Norouzi}]{saharia2022image}
\bibinfo{author}{C.~Saharia}, \bibinfo{author}{J.~Ho},
  \bibinfo{author}{W.~Chan}, \bibinfo{author}{T.~Salimans},
  \bibinfo{author}{D.~J. Fleet}, \bibinfo{author}{M.~Norouzi},
\newblock \bibinfo{title}{Image super-resolution via iterative refinement},
\newblock \bibinfo{journal}{IEEE transactions on pattern analysis and machine
  intelligence} \bibinfo{volume}{45} (\bibinfo{year}{2022})
  \bibinfo{pages}{4713--4726}.
\bibitem[{Anantatamukala et~al.(2023)Anantatamukala, Krishna, and
  Dahotre}]{anantatamukala2023generative}
\bibinfo{author}{A.~Anantatamukala}, \bibinfo{author}{K.~M. Krishna},
  \bibinfo{author}{N.~B. Dahotre},
\newblock \bibinfo{title}{Generative adversarial networks assisted machine
  learning based automated quantification of grain size from scanning electron
  microscope back scatter images},
\newblock \bibinfo{journal}{Materials Characterization} \bibinfo{volume}{206}
  (\bibinfo{year}{2023}) \bibinfo{pages}{113396}.
\bibitem[{Shahriar(2022)}]{shahriar2022gan}
\bibinfo{author}{S.~Shahriar},
\newblock \bibinfo{title}{Gan computers generate arts? a survey on visual arts,
  music, and literary text generation using generative adversarial network},
\newblock \bibinfo{journal}{Displays} \bibinfo{volume}{73}
  (\bibinfo{year}{2022}) \bibinfo{pages}{102237}.
\bibitem[{Ledig et~al.(2017)Ledig, Theis, Husz{\'a}r, Caballero, Cunningham,
  Acosta, Aitken, Tejani, Totz, Wang et~al.}]{ledig2017photo}
\bibinfo{author}{C.~Ledig}, \bibinfo{author}{L.~Theis},
  \bibinfo{author}{F.~Husz{\'a}r}, \bibinfo{author}{J.~Caballero},
  \bibinfo{author}{A.~Cunningham}, \bibinfo{author}{A.~Acosta},
  \bibinfo{author}{A.~Aitken}, \bibinfo{author}{A.~Tejani},
  \bibinfo{author}{J.~Totz}, \bibinfo{author}{Z.~Wang}, et~al.,
\newblock \bibinfo{title}{Photo-realistic single image super-resolution using a
  generative adversarial network},
\newblock in: \bibinfo{booktitle}{Proceedings of the IEEE conference on
  computer vision and pattern recognition}, \bibinfo{year}{2017}, pp.
  \bibinfo{pages}{4681--4690}.
\bibitem[{Wang et~al.(2018)Wang, Yu, Wu, Gu, Liu, Dong, Qiao, and
  Change~Loy}]{wang2018esrgan}
\bibinfo{author}{X.~Wang}, \bibinfo{author}{K.~Yu}, \bibinfo{author}{S.~Wu},
  \bibinfo{author}{J.~Gu}, \bibinfo{author}{Y.~Liu}, \bibinfo{author}{C.~Dong},
  \bibinfo{author}{Y.~Qiao}, \bibinfo{author}{C.~Change~Loy},
\newblock \bibinfo{title}{Esrgan: Enhanced super-resolution generative
  adversarial networks},
\newblock in: \bibinfo{booktitle}{Proceedings of the European conference on
  computer vision (ECCV) workshops}, \bibinfo{year}{2018}, pp.
  \bibinfo{pages}{0--0}.
\bibitem[{Zhang et~al.(2018)Zhang, Li, and Yu}]{zhang2018convergence}
\bibinfo{author}{Z.~Zhang}, \bibinfo{author}{M.~Li}, \bibinfo{author}{J.~Yu},
\newblock \bibinfo{title}{On the convergence and mode collapse of gan},
\newblock in: \bibinfo{booktitle}{SIGGRAPH Asia 2018 Technical Briefs},
  \bibinfo{year}{2018}, pp. \bibinfo{pages}{1--4}.
\bibitem[{Rombach et~al.(2022)Rombach, Blattmann, Lorenz, Esser, and
  Ommer}]{rombach2022high}
\bibinfo{author}{R.~Rombach}, \bibinfo{author}{A.~Blattmann},
  \bibinfo{author}{D.~Lorenz}, \bibinfo{author}{P.~Esser},
  \bibinfo{author}{B.~Ommer},
\newblock \bibinfo{title}{High-resolution image synthesis with latent diffusion
  models},
\newblock in: \bibinfo{booktitle}{Proceedings of the IEEE/CVF conference on
  computer vision and pattern recognition}, \bibinfo{year}{2022}, pp.
  \bibinfo{pages}{10684--10695}.
\bibitem[{Kawar et~al.(2022)Kawar, Elad, Ermon, and Song}]{kawar2022denoising}
\bibinfo{author}{B.~Kawar}, \bibinfo{author}{M.~Elad},
  \bibinfo{author}{S.~Ermon}, \bibinfo{author}{J.~Song},
\newblock \bibinfo{title}{Denoising diffusion restoration models},
\newblock \bibinfo{journal}{Advances in Neural Information Processing Systems}
  \bibinfo{volume}{35} (\bibinfo{year}{2022}) \bibinfo{pages}{23593--23606}.
\bibitem[{Song et~al.(2020)Song, Meng, and Ermon}]{song2020denoising}
\bibinfo{author}{J.~Song}, \bibinfo{author}{C.~Meng},
  \bibinfo{author}{S.~Ermon},
\newblock \bibinfo{title}{Denoising diffusion implicit models},
\newblock \bibinfo{journal}{arXiv preprint arXiv:2010.02502}
  (\bibinfo{year}{2020}).
\bibitem[{Ulhaq et~al.(2022)Ulhaq, Akhtar, and Pogrebna}]{ulhaq2022efficient}
\bibinfo{author}{A.~Ulhaq}, \bibinfo{author}{N.~Akhtar},
  \bibinfo{author}{G.~Pogrebna},
\newblock \bibinfo{title}{Efficient diffusion models for vision: A survey},
\newblock \bibinfo{journal}{arXiv preprint arXiv:2210.09292}
  (\bibinfo{year}{2022}).
\bibitem[{Falaleev and Orlov(2025)}]{falaleev2025self}
\bibinfo{author}{N.~Falaleev}, \bibinfo{author}{N.~Orlov},
\newblock \bibinfo{title}{Self-controlled diffusion for denoising in scientific
  imaging},
\newblock \bibinfo{journal}{arXiv preprint arXiv:2504.16951}
  (\bibinfo{year}{2025}).
\bibitem[{Saharia et~al.(2022)Saharia, Chan, Chang, Lee, Ho, Salimans, Fleet,
  and Norouzi}]{saharia2022palette}
\bibinfo{author}{C.~Saharia}, \bibinfo{author}{W.~Chan},
  \bibinfo{author}{H.~Chang}, \bibinfo{author}{C.~Lee},
  \bibinfo{author}{J.~Ho}, \bibinfo{author}{T.~Salimans},
  \bibinfo{author}{D.~Fleet}, \bibinfo{author}{M.~Norouzi},
\newblock \bibinfo{title}{Palette: Image-to-image diffusion models},
\newblock in: \bibinfo{booktitle}{ACM SIGGRAPH 2022 conference proceedings},
  \bibinfo{year}{2022}, pp. \bibinfo{pages}{1--10}.
\bibitem[{Mandal et~al.(2025)Mandal, Chattopadhyay, Tong, and
  Chakravarthula}]{mandal2025unicorn}
\bibinfo{author}{D.~Mandal}, \bibinfo{author}{S.~Chattopadhyay},
  \bibinfo{author}{G.~Tong}, \bibinfo{author}{P.~Chakravarthula},
\newblock \bibinfo{title}{Unicorn: Latent diffusion-based unified controllable
  image restoration network across multiple degradations},
\newblock \bibinfo{journal}{arXiv preprint arXiv:2503.15868}
  (\bibinfo{year}{2025}).
\bibitem[{Bachmann et~al.(2010)Bachmann, Hielscher, and
  Schaeben}]{bachmann2010texture}
\bibinfo{author}{F.~Bachmann}, \bibinfo{author}{R.~Hielscher},
  \bibinfo{author}{H.~Schaeben},
\newblock \bibinfo{title}{Texture analysis with mtex--free and open source
  software toolbox},
\newblock \bibinfo{journal}{Solid state phenomena} \bibinfo{volume}{160}
  (\bibinfo{year}{2010}) \bibinfo{pages}{63--68}.
\bibitem[{Batzolis et~al.(2021)Batzolis, Stanczuk, Sch{\"o}nlieb, and
  Etmann}]{batzolis2021conditional}
\bibinfo{author}{G.~Batzolis}, \bibinfo{author}{J.~Stanczuk},
  \bibinfo{author}{C.-B. Sch{\"o}nlieb}, \bibinfo{author}{C.~Etmann},
\newblock \bibinfo{title}{Conditional image generation with score-based
  diffusion models},
\newblock \bibinfo{journal}{arXiv preprint arXiv:2111.13606}
  (\bibinfo{year}{2021}).
\bibitem[{Lin et~al.(2022)Lin, Cheng, Wu, and Shen}]{lin2022cat}
\bibinfo{author}{H.~Lin}, \bibinfo{author}{X.~Cheng}, \bibinfo{author}{X.~Wu},
  \bibinfo{author}{D.~Shen},
\newblock \bibinfo{title}{Cat: Cross attention in vision transformer},
\newblock in: \bibinfo{booktitle}{2022 IEEE international conference on
  multimedia and expo (ICME)}, \bibinfo{organization}{IEEE},
  \bibinfo{year}{2022}, pp. \bibinfo{pages}{1--6}.
\bibitem[{Perez et~al.(2018)Perez, Strub, De~Vries, Dumoulin, and
  Courville}]{perez2018film}
\bibinfo{author}{E.~Perez}, \bibinfo{author}{F.~Strub},
  \bibinfo{author}{H.~De~Vries}, \bibinfo{author}{V.~Dumoulin},
  \bibinfo{author}{A.~Courville},
\newblock \bibinfo{title}{Film: Visual reasoning with a general conditioning
  layer},
\newblock in: \bibinfo{booktitle}{Proceedings of the AAAI conference on
  artificial intelligence}, volume~\bibinfo{volume}{32}, \bibinfo{year}{2018}.
\bibitem[{Ronneberger et~al.(2015)Ronneberger, Fischer, and
  Brox}]{ronneberger2015u}
\bibinfo{author}{O.~Ronneberger}, \bibinfo{author}{P.~Fischer},
  \bibinfo{author}{T.~Brox},
\newblock \bibinfo{title}{U-net: Convolutional networks for biomedical image
  segmentation},
\newblock in: \bibinfo{booktitle}{Medical image computing and computer-assisted
  intervention--MICCAI 2015: 18th international conference, Munich, Germany,
  October 5-9, 2015, proceedings, part III 18},
  \bibinfo{organization}{Springer}, \bibinfo{year}{2015}, pp.
  \bibinfo{pages}{234--241}.
\bibitem[{Salimans et~al.(2017)Salimans, Karpathy, Chen, and
  Kingma}]{salimans2017pixelcnn++}
\bibinfo{author}{T.~Salimans}, \bibinfo{author}{A.~Karpathy},
  \bibinfo{author}{X.~Chen}, \bibinfo{author}{D.~P. Kingma},
\newblock \bibinfo{title}{Pixelcnn++: Improving the pixelcnn with discretized
  logistic mixture likelihood and other modifications},
\newblock \bibinfo{journal}{arXiv preprint arXiv:1701.05517}
  (\bibinfo{year}{2017}).
\bibitem[{He et~al.(2016)He, Zhang, Ren, and Sun}]{he2016deep}
\bibinfo{author}{K.~He}, \bibinfo{author}{X.~Zhang}, \bibinfo{author}{S.~Ren},
  \bibinfo{author}{J.~Sun},
\newblock \bibinfo{title}{Deep residual learning for image recognition}
  (\bibinfo{year}{2016}) \bibinfo{pages}{770--778}.
\bibitem[{Hendrycks and Gimpel(2016)}]{hendrycks2016gaussian}
\bibinfo{author}{D.~Hendrycks}, \bibinfo{author}{K.~Gimpel},
\newblock \bibinfo{title}{Gaussian error linear units (gelus)},
\newblock \bibinfo{journal}{arXiv preprint arXiv:1606.08415}
  (\bibinfo{year}{2016}).
\bibitem[{Loshchilov and Hutter(2017)}]{loshchilov2017decoupled}
\bibinfo{author}{I.~Loshchilov}, \bibinfo{author}{F.~Hutter},
\newblock \bibinfo{title}{Decoupled weight decay regularization},
\newblock \bibinfo{journal}{arXiv preprint arXiv:1711.05101}
  (\bibinfo{year}{2017}).
\bibitem[{Ånes et~al.(2024)Ånes, Lervik, Natlandsmyr, Bergh, Prestat, Bugten,
  Østvold, Xu, Francis, and Nord}]{hakon_wiik_anes_2024_11432173}
\bibinfo{author}{H.~W. Ånes}, \bibinfo{author}{L.~A.~H. Lervik},
  \bibinfo{author}{O.~Natlandsmyr}, \bibinfo{author}{T.~Bergh},
  \bibinfo{author}{E.~Prestat}, \bibinfo{author}{A.~V. Bugten},
  \bibinfo{author}{E.~M. Østvold}, \bibinfo{author}{Z.~Xu},
  \bibinfo{author}{C.~Francis}, \bibinfo{author}{M.~Nord},
  \bibinfo{title}{pyxem/kikuchipy: kikuchipy 0.10.0}, \bibinfo{year}{2024}.
  \URLprefix \url{https://doi.org/10.5281/zenodo.11432173}.
  \DOIprefix\doi{10.5281/zenodo.11432173}.
\bibitem[{Field(1997)}]{field1997recent}
\bibinfo{author}{D.~P. Field},
\newblock \bibinfo{title}{Recent advances in the application of orientation
  imaging},
\newblock \bibinfo{journal}{Ultramicroscopy} \bibinfo{volume}{67}
  (\bibinfo{year}{1997}) \bibinfo{pages}{1--9}.

\end{thebibliography}
	


\end{document}